%% file: boogaard-miri-hudf-alma.tex
\documentclass[twocolumn,times,twocolappendix]{aastex631}

\usepackage[]{newtxtext}
\usepackage[]{newtxmath}

\usepackage{boogaard-commands}

\received{August 31, 2023}
\revised{February 20, 2024}
\accepted{April 22, 2024}

\shorttitle{JWST/MIRI Stellar Structure of ALMA-selected Galaxies in the HUDF}
\shortauthors{Boogaard et al.}

\graphicspath{{./}{figures/}}

\newcommand{\MPIA}{\affiliation{Max Planck Institute for Astronomy, K\"onigstuhl 17,  69117 Heidelberg, Germany}}

\newcommand{\DAWN}{\affiliation{Cosmic Dawn Center (DAWN), Denmark}}

\newcommand{\DTU}{\affiliation{DTU Space, Technical University of Denmark, Elektrovej, Building 328, 2800, Kgs. Lyngby, Denmark}}

\newcommand{\DARK}{\affil{DARK, Niels Bohr Institute, University of Copenhagen, Jagtvej 128, 2200 Copenhagen, Denmark}}

\newcommand{\Stockholm}{\affiliation{Department of Astronomy, Stockholm University, Oscar Klein Centre, AlbaNova University Centre, 106 91 Stockholm, Sweden}}

\newcommand{\NRAO}{\affiliation{National Radio Astronomy Observatory, Pete V. Domenici Array Science Center, P.O. Box O, Socorro, NM 87801, USA}}

\newcommand{\CABcda}{\affiliation{Centro de Astrobiolog\'{i}a (CAB), CSIC-INTA, Ctra. de Ajalvir km 4, Torrej\'{o}n de Ardoz, E-28850, Madrid, Spain}}
\newcommand{\CABcbdc}{\affiliation{Centro de Astrobiolog\'{i}a (CAB), CSIC-INTA, Camino Bajo del Castillo s/n, 28692 Villanueva de la Ca\~{n}ada, Madrid, Spain}}

\newcommand{\ESAC}{\affiliation{Telespazio UK for the European Space Agency, ESAC, Camino Bajo del Castillo s/n, 28692 Villanueva de la Ca\~nada, Spain}}

\newcommand{\Kapteyn}{\affiliation{Kapteyn Astronomical Institute, University of Groningen, P.O. Box 800, 9700AV Groningen, The Netherlands}}

\newcommand{\Leiden}{\affil{Leiden Observatory,  Leiden University, PO Box 9513, NL-2300 RA Leiden, The Netherlands}}

\newcommand{\UCL}{\affiliation{Dept. of Physics and Astronomy, University College London, Gower Street, London WC1E 6BT, United Kingdom}}

\newcommand{\Edinburgh}{\affiliation{UK Astronomy Technology Centre, Royal Observatory Edinburgh, Blackford Hill, Edinburgh EH9 3HJ, UK}}

\newcommand{\ITA}{\affil{Institute for Theoretical Physics, Heidelberg University, Philosophenweg 12, D–69120, Heidelberg, Germany}}

\newcommand{\Cologne}{\affil{I.Physikalisches Institut der Universit\"{a}t zu K\"{o}ln, Z\"{u}lpicher Str. 77, 50937 K\"{o}ln, Germany}}

\newcommand{\MPIfR}{\affil{Max Planck Institute for Radiosastronomie, Auf dem H\"{u}gel 69, 53121 Bonn, Germany}}

\newcommand{\Geneva}{\affil{Departement d'Astronomie, University of Geneva, Chemin Pegasi 51, 1290 Versoix, Switzerland}}

\newcommand{\Marseille}{\affil{Aix Marseille Univ, CNRS, CNES, LAM, Marseille, France}}

\newcommand{\SPL}{\affil{School of Physics \& Astronomy, Space Research Centre, Space Park Leicester, University of Leicester, 92 Corporation Road, Leicester LE4 5SP, UK}}

\newcommand{\Dublin}{\affil{Dublin Institute for Advanced Studies, Astronomy \& Astrophysics Section, 31 Fitzwilliam Place, Dublin 2, Ireland}}
 
\begin{document}

\title{MIDIS: JWST/MIRI reveals the Stellar Structure of ALMA-selected Galaxies in the Hubble--UDF at Cosmic Noon}

\correspondingauthor{Leindert A. Boogaard}
\email{boogaard@mpia.de}

\author[0000-0002-3952-8588]{Leindert A. Boogaard} \MPIA

\author[0000-0001-9885-4589]{Steven Gillman} \DAWN \DTU

\author[0000-0003-0470-8754]{Jens Melinder} \Stockholm

\author[0000-0003-4793-7880]{Fabian Walter} \MPIA \NRAO

\author[0000-0002-9090-4227]{Luis Colina} \CABcda

\author[0000-0002-3005-1349]{G\"{o}ran \"{O}stlin} \Stockholm

\author[0000-0001-8183-1460]{Karina I. Caputi} \Kapteyn

\author[0000-0001-8386-3546]{Edoardo Iani} \Kapteyn

\author[0000-0003-4528-5639]{Pablo P\'{e}rez-Gonz\'{a}lez} \CABcda

\author[0000-0001-5434-5942]{Paul van der Werf} \Leiden

\author[0000-0002-2554-1837]{Thomas R.~Greve} \DAWN \DTU \UCL

\author[0000-0001-7416-7936]{Gillian Wright} \Edinburgh

\author[0000-0001-6794-2519]{Almudena Alonso-Herrero} \CABcbdc

\author{Javier \'Alvarez-M\'arquez} \CABcda

\author{Marianna Annunziatella} \CABcda

\author{Arjan Bik} \Stockholm

\author[0000-0001-8582-7012]{Sarah Bosman} \MPIA \ITA

\author[0000-0001-6820-0015]{Luca Costantin}  \CABcda

\author[0000-0003-2119-277X]{Alejandro Crespo G\'{o}mez} \CABcda

\author[0000-0003-0589-5969]{Dan Dicken} \Edinburgh

\author{Andreas Eckart} \Cologne \MPIfR

\author[0000-0002-4571-2306]{Jens Hjorth} \DARK

\author[0000-0002-2624-1641]{Iris Jermann} \DAWN \DTU

\author[0000-0002-0690-8824]{Alvaro Labiano} \ESAC \CABcbdc

\author[0000-0001-5710-8395]{Danial Langeroodi} \DARK

\author[0000-0001-5492-4522]{Romain A. Meyer} \Geneva

\author[0000-0002-3305-9901]{Thibaud Moutard} \Marseille

\author{Florian Pei{\ss}ker} \Cologne

\author[0000-0002-0932-4330]{John P. Pye} \SPL

\author[0000-0002-5104-8245]{Pierluigi Rinaldi} \Kapteyn

\author{Tuomo V. Tikkanen} \SPL

\author{Martin Topinka} \Dublin

\author[0000-0002-1493-300X]{Thomas Henning} \MPIA

\begin{abstract}
  We present deep James Webb Space Telescope (JWST)/MIRI F560W
  observations of a flux-limited, ALMA-selected sample of 28 galaxies
  at $z=0.5$--3.7 in the Hubble Ultra Deep Field (HUDF).  The data
  from the MIRI Deep Imaging Survey (MIDIS) reveal the stellar
  structure of the HUDF galaxies at rest-wavelengths of
  $\lambda>1$\,\micron\ for the first time.  We revise the stellar
  mass estimates using new JWST photometry and find good agreement
  with pre-JWST analysis; the few discrepancies can be explained by
  blending issues in the earlier lower-resolution Spitzer data.  At
  $z\sim2.5$, the resolved rest-frame near-infrared (1.6\,\micron)
  structure of the galaxies is significantly more smooth and centrally
  concentrated than seen by HST at rest-frame 450\,nm (F160W), with
  effective radii of $\Remiri=1$--5\,kpc and S\'{e}rsic indices mostly
  close to an exponential (disk-like) profile ($n\approx1$), up to
  $n\approx5$ (excluding AGN).  We find an average size ratio of
  $\Remiri/\Rehst\approx0.7$ that decreases with stellar mass. The
  stellar structure of the ALMA-selected galaxies is indistinguishable
  from a HUDF reference sample of all galaxies with a MIRI flux
  density greater than 1\,$\mu$Jy.  We supplement our analysis with
  custom-made, position-dependent, empirical PSF models for the F560W
  observations. The results imply that a smoother stellar structure is
  in place in massive gas-rich, star-forming galaxies at Cosmic Noon,
  despite a more clumpy rest-frame optical appearance, placing
  additional constraints on galaxy formation simulations.  As a next
  step, matched-resolution, resolved ALMA observations will be crucial
  to further link the mass- and light-weighted galaxy structures to
  the dusty interstellar medium.
\end{abstract}

\keywords{Galaxy structure(622), High-redshift galaxies (734), Galaxy
  Evolution (594), James Webb Space Telescope(2291), Millimeter
  astronomy(1061)}

\section{Introduction} \label{sec:intro}
\begin{figure*}[t]
  \includegraphics[width=\textwidth]{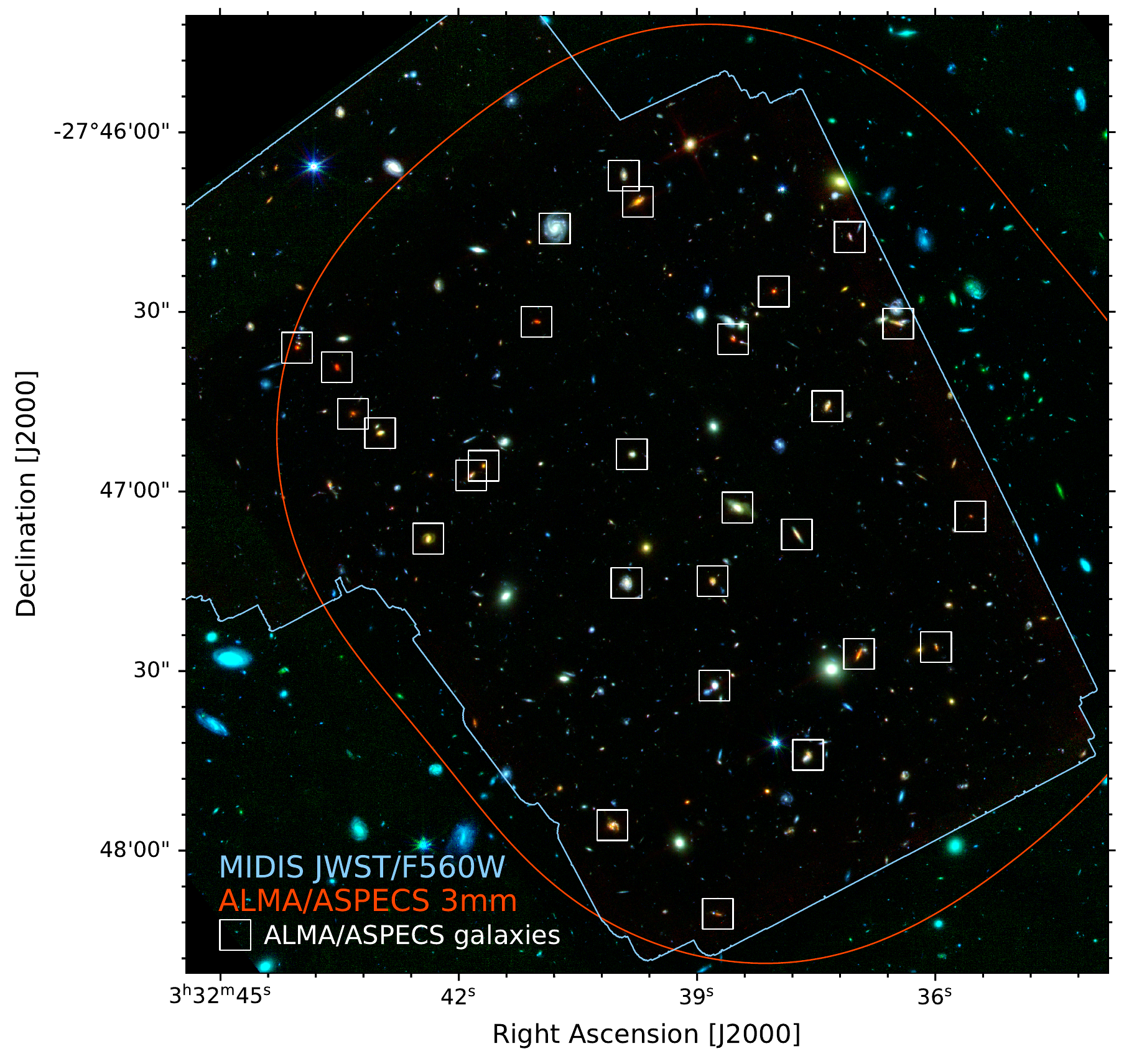}
  \caption{MIRI/F560W, NIRCam/F182M and HST/F814W (RGB) image
    highlighting the footprints of the MIRI Deep Imaging Survey
    (MIDIS) and the ALMA Spectroscopic Survey in the Hubble Ultra Deep
    Field (ASPECS; at 3mm, which encompasses the 1.2\,mm footprint).
    The squares indicate the ALMA-selected galaxies, shown close-up in
    \autoref{fig:cutouts}. \label{fig:midis-aspecs-field}}
\end{figure*}

Galaxy formation reached its high point around 10\,billion years ago
during the peak of cosmic star formation at Cosmic Noon
($z\approx1-3$; \citealt{Madau2014}).  The typical star-forming
galaxy, located on the galaxy main sequence (MS), formed around
$8\times$ more stars during this epoch than a galaxy with similar
stellar mass in the local universe \citep[e.g.,][]{Whitaker2014}.
This rise in the global star formation rates goes hand-in-hand with an
increase in the global molecular gas content of galaxies
\citep{Walter2014, Walter2016, Riechers2019, Decarli2020,
  Boogaard2023}.  Galaxies at Cosmic Noon are significantly more
gas-rich than their local counterparts at fixed stellar mass, with a
total cold gas mass that can exceed the total mass in stars
($M_{\rm mol} / M_{*} \geq 1$; e.g., \citealt{Tacconi2013,
  Tacconi2018}).  These large gas fractions are expected to have a
significant impact on the conditions inside the interstellar medium
(ISM), potentially leading to the apparently clumpy star-forming
structures in the rest-frame UV/optical and increased ionized-gas
velocity dispersions (see \citealt{ForsterSchreiber2020} for a
review).

Yet, the older stellar populations---best traced at rest-frame
wavelengths $\lambda>1$\,\micron---which contain the bulk of the
stellar mass, have remained unresolved due to the relatively large
point-spread-function (PSF) of the Spitzer/IRAC instrument
\citep[1\farcs6--2\farcs0, or 13--17\,kpc at $z\approx2$;][]{Fazio2004} and their flux has remained uncertain due to the
complex deblending of neighbouring sources
\citep[e.g.,][]{Labbe2015}.  As a result, the total stellar mass and
structure of the galaxies remain uncertain, as well as the gas
fraction and its potential impact on galaxy structure.  This leaves
open important questions, such as whether the galaxies are
intrinsically clumpy, or whether their underlying stellar-mass
distribution is smoother.

The Mid-Infrared Instrument (MIRI) on board the James Webb Space
Telescope (JWST) can now provide sensitive, high-resolution imaging at
wavelengths of 5.6\,\micron\ and above \citep{Wright2023}.  This allows
one to trace the rest-frame near-infrared light of galaxies at
$z\approx2.5$ ($\lambda_{\rm rest} \approx 1.6$\,\micron) and beyond
for the first time, with unprecedented spatial resolution
($\approx 0\farcs2$ at 5.6\,\micron, or 1.7\,kpc at $z\approx2$).

In this paper, we utilise deep JWST/MIRI observations to study a flux-limited
sample of dust continuum- and cold gas-selected galaxies in the Hubble
Ultra Deep Field (HUDF).  The sample is taken from the ALMA Spectroscopic
Survey (ASPECS) Large Program \citep{Walter2016, Decarli2019,
  Decarli2020} that performed flux-limited spectral scans in the
1.2\,mm and 3\,mm bands to detect molecular gas via $^{12}$CO across
cosmic time \citep{Gonzalez-Lopez2019, Boogaard2019} while
simultaneously obtaining extremely sensitive 1.2\,mm dust continuum
imaging over the HUDF \citep[][]{Gonzalez-Lopez2020, Aravena2020}.

This paper is organised as follows: in \autoref{sec:observations} we
present the JWST observations, including NIRCam imaging and
slitless spectroscopy. In \autoref{sec:analysis} we discuss the
updated properties of the full ALMA sample in the JWST era. We then
analyse the rest-frame near-infrared morphology focusing on the
$z\ge2$ galaxies, now probed by the deep MIRI/F560W observations, in
\autoref{sec:results} and discuss the results in
\autoref{sec:discussion}.  We adopt a \cite{Chabrier2003} initial mass
function and a \cite{PlanckCollaboration2018a} cosmology (flat
$\Lambda$CDM with $H_{0} = 67.7$\,km\,s$^{-1}$\,Mpc$^{-1}$,
$\Omega_{m} = 0.31$ and $\Omega_{\Lambda} = 0.69$).  We use $\log$ to
denote $\log_{10}$ and $\ln$ for the natural logarithm.  We report
magnitudes in the AB system \citep{Oke1983}.

\section{JWST Observations}
\label{sec:observations}
\begin{figure*}[t]
  \includegraphics[width=\textwidth]{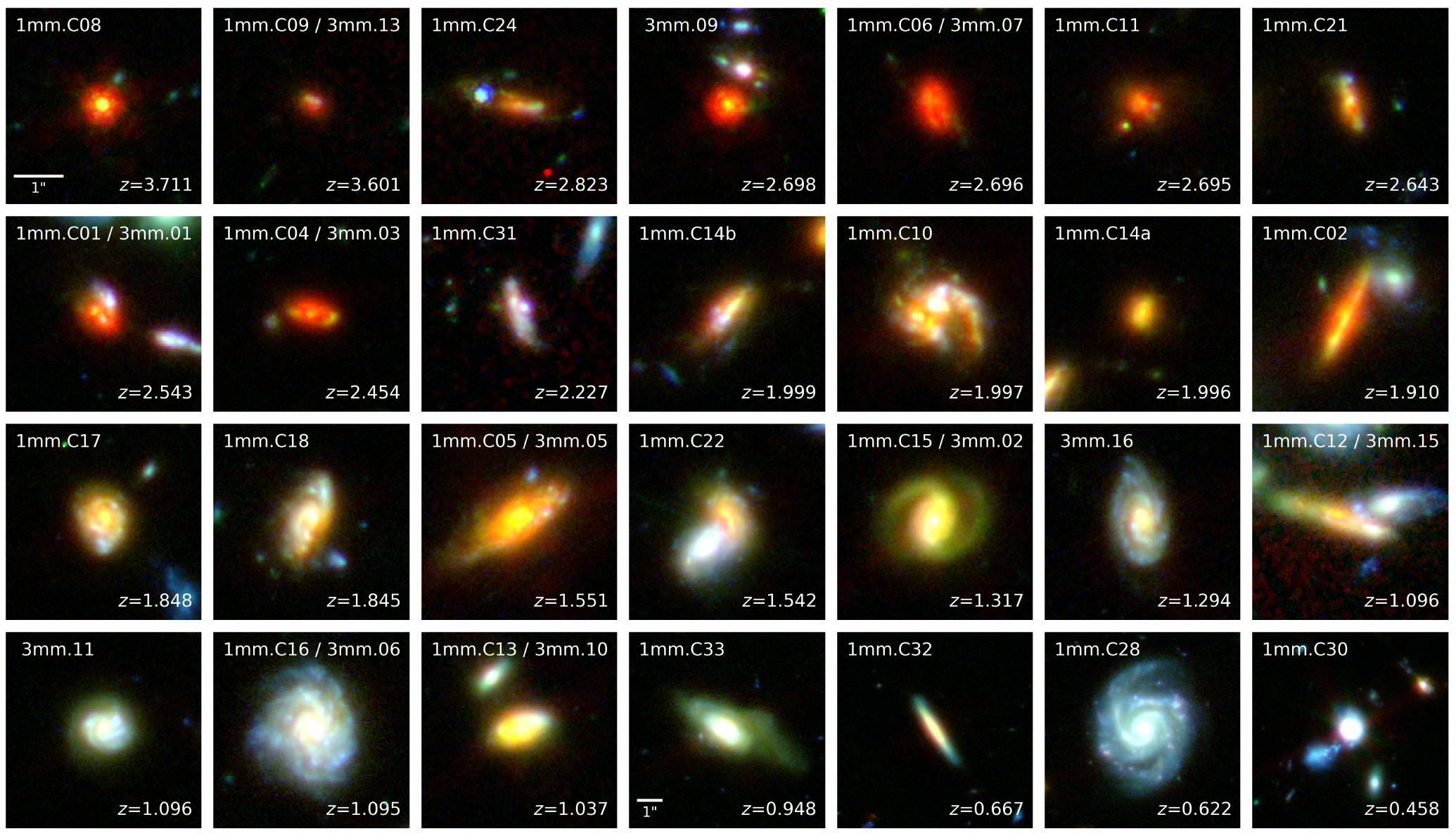}
  \caption{MIRI/F560W, NIRCam/F182M, HST/F814W (RGB) cutouts of the
    flux-limited ALMA/ASPECS sample in the MIRI Deep Imaging Survey
    footprint.  The cutouts are ordered by decreasing redshifts and
    are $4\arcsec\times4\arcsec$, except the last four galaxies at
    $z<1$, which are $8\arcsec\times8\arcsec$ (as indicated by the
    scalebars). See \autoref{tab:sources} for more information on the
    source properties.\label{fig:cutouts}}
\end{figure*}
The MIRI Deep Imaging Survey (MIDIS) is a deep JWST/MIRI survey of the
Hubble UDF conducted by the MIRI European Consortium GTO program
(Prog. ID 1283, PI: G. \"{Ostlin}).  MIDIS was designed as a single
F560W deep field, spread over 6 observations totaling 63\,hours
(48.8\,h net exposure time), with 100 groups, 100 integrations and a
10-point random cycling dither pattern.  The majority of the MIDIS
observations (4/6) were executed between 2--6 December 2022, while the
fifth observation was executed on 20 December 2022, at a slightly
different central position and position angle, due to several
safe-mode incidents of the observatory in the meantime.  The data were
reduced using the official JWST pipeline version 1.12.3 (CRDS pmap
1137; \citealt{jwst-pipeline-1123}).  In addition to the default
pipeline steps, we perform custom routines to deal with cosmic
showers, background variations and other well known MIRI instrument
effects as well as the absolute and relative astrometric alignment of
the exposures. These additional steps use a similar approach as those
taken in other high-redshift studies
\citep[e.g.][]{Bagley2023a,Iani2022, Rinaldi2023}.  The final mosaic
consists of a total of 96 exposures with 50 unique dithers and has a
total exposure time of 41.4\,h, reaching a $5\sigma$ point-source
depth of $28.6$ mag (accounting for correlated noise), and covers a
total area of 4.7\,arcmin$^{2}$. Further details will be described in
\"{O}stlin et al. (in prep.).

In addition to the MIRI data, we use the publicly available
JWST/NIRCam imaging over the HUDF, taken as part of the programs JADES
\citep[Prog. ID 1180;][]{Rieke2023}, JEMS \citep[Prog. ID
1963;][]{Williams2023} and FRESCO \citep[Prog. ID 1895;][]{Oesch2023}.
We use the combined medium-band and wide-band observations as released
by the JADES team (Data Release 1) and the \textsc{Grizli} Image
Release v7.0, July
2023\footnote{\url{https://dawn-cph.github.io/dja/imaging/v7/}}.

JWST/NIRCam slitless spectroscopy over the HUDF was taken as part of
the FRESCO program and used to obtain redshift information of galaxies
without prior secure redshift information.  These observations were
taken in the F444W filter (covering 3.9--5.0\micron, limited to 4.4
\micron\ for part of the survey area) and a single grism orientation
(GrismR) with a resolution of $R\sim1600$.  Two hours of exposure time
is obtained in 8 exposures, designed to reach a $5\sigma$ point-source
depth for an unresolved emission line of
$2\times10^{-18}$\,erg\,s$^{-1}$\,cm$^{-2}$.  For more details on the
spectral coverage and extraction we refer to \cite{Oesch2023}.

\section{Analysis}
\label{sec:analysis}

\subsection{Sample}
\label{sec:sample}

The flux-limited ASPECS sample consists of all galaxies that are
detected in the deep 1.2\,mm dust continuum image from ASPECS, that
has an unprecedented 9.3\,$\mu$Jy\,beam$^{-1}$ rms sensitivity and a
resolution (beam size) of roughly 1\farcs3
\citep[][]{Gonzalez-Lopez2020, Aravena2020}.  From this sample, we
discard three sources (1mm.C27, 1mm.C29, and 1mm.C34) that remain
without counterparts in the optical-FIR and also the new JWST imaging
and are potentially false-positives, consistent with the fidelity
estimates of the sample \citep[][]{Gonzalez-Lopez2020}.  The
continuum-selected sample includes all of the $^{12}$CO-selected
galaxies \citep{Gonzalez-Lopez2019, Boogaard2019}, as discussed in
\cite{Boogaard2020}, apart from two sources detected in $^{12}$CO
only, which are included in \autoref{tab:sources} for completeness,
but excluded from the analysis where relevant (one of these sources is
part of the continuum-faint sample, see
\autoref{sec:struct-param-z14}).  The sample also encompasses all
1\,mm continuum sources from shallower ALMA data in the HUDF area
\citep[e.g.,][]{Dunlop2017, Franco2018, Hatsukade2018}.

As the Rayleigh-Jeans tail of the dust continuum
emission is nearly always optically thin, the
1.2\,mm selection is essentially a cold dust mass selection (the phase
which contains most of the dust mass) and effectively also a cold,
molecular gas mass selection (e.g., \citealt{Hildebrand1983,
  Scoville2014, Scoville2016}; see \citealt{Aravena2020}, their Appendix A,
on the consistency between the dust- and $^{12}$CO-based molecular gas
mass estimates).

The complete sample consists of 35 galaxies, 28 of which are inside of
the MIRI/F560W footprint in the HUDF, as shown in
\autoref{fig:midis-aspecs-field} (see also \autoref{tab:sources}), and
span a redshift range of $z=0.45$--3.71.  We show MIRI/NIRCam/HST
cutouts in \autoref{fig:cutouts}.

\subsection{Redshifts}
\label{sec:redshifts}
Most ASPECS sources already have spectroscopic redshifts from the deep
MUSE HUDF and MXDF Surveys \citep[][see \citealt{Boogaard2019} and
\citealt{Aravena2020}]{Bacon2017, Bacon2023} and/or their multi-$J$ CO
and \CI\ emission \citep{Boogaard2020}.  The few sources with missing
spectroscopic redshifts lie primarily between $z=1.5$ and $2.9$ where
there are no bright emission lines covered by MUSE, including
$1.74 < z < 2.0$, where there is also no low-$J$ CO-coverage from
ASPECS (see \citealt{Boogaard2019}, Fig.~1).  This gap in
spectroscopic redshift coverage is now mostly filled by FRESCO, that
covers a range of IR emission lines (including \HeiIR, \FeiiIRa,
\Pagamma\ and \Pabeta\ lines between $2<z<3.6$, \Paalpha\ between
$1<z<1.6$, and many fainter lines).

We extract FRESCO spectra for all ALMA sources, removing the continuum
using a median filter, and search for bright emission lines in a
$\Delta z/(1+z) = 0.1$ window around the known redshift.  We detect
line emission in nearly all sources that have spectral coverage of the
brightest emission lines.  Notably, we confirm the earlier photometric
redshifts for all sources \citep[see][]{Aravena2020}.  We detect
\Siii\ in the highest redshift sources and confirm the tentative MUSE
redshifts for two known blended sources, 1mm.C08 and 1mm.C22 (which
corresponds to the northern galaxy).  For 1mm.C2 no lines are covered
in the FRESCO wavelength range, but we report a spectroscopic redshift
of $z=1.91$ based on the detection of \Ha\ in the NGDEEP slitless
spectroscopy \citep{Bagley2023b,Pirzkal2023}.  The updated redshift
information is given in \autoref{tab:sources}.  The FRESCO spectra of
the aforementioned sources are shown in \autoref{fig:spectra} in
\autoref{sec:fresco-spectra}.  Notably, all sources in the sample now
have a spectroscopic redshift.

\subsection{Spectral energy distributions}
\label{sec:spectr-energy-distr}
\begin{figure}[t]
  \centering
  \includegraphics[height=0.3\textheight]{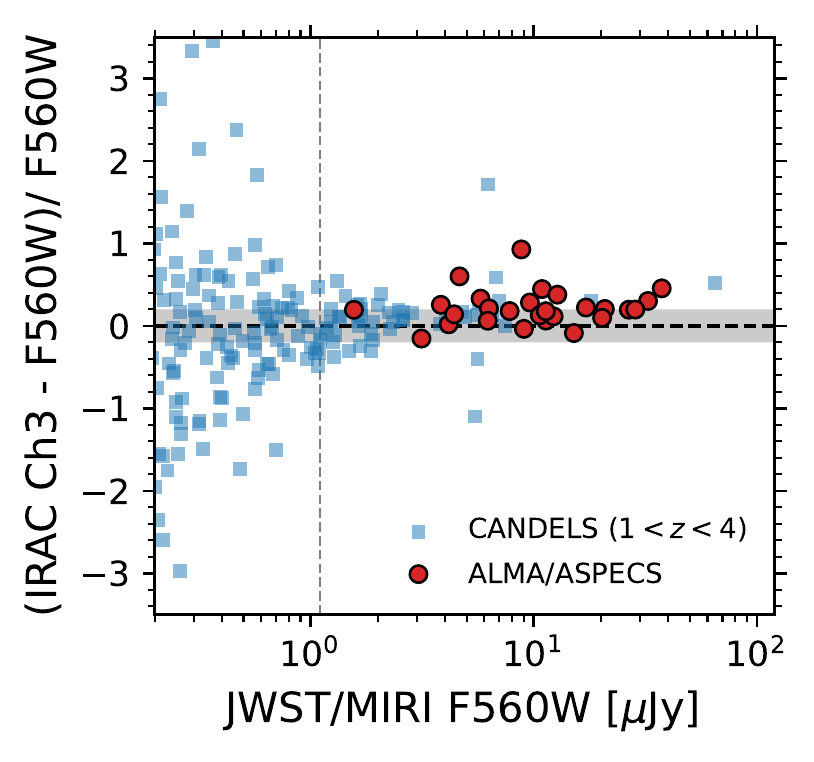}
\caption{JWST/MIRI F560W fluxes compared to the Spitzer/IRAC Channel 3 (5.8\micron) fluxes from CANDELS \citep{Guo2013} for galaxies at $1<z<4$.  The vertical gray dashed line shows the Spitzer $5\sigma$ flux limit.  The ALMA/ASPECS sources are shown in red. The grey band shows a $20\%$ relative flux difference.
  \label{fig:miri-spitzer}}
\end{figure}

\begin{figure*}[t]
\includegraphics[width=\textwidth]{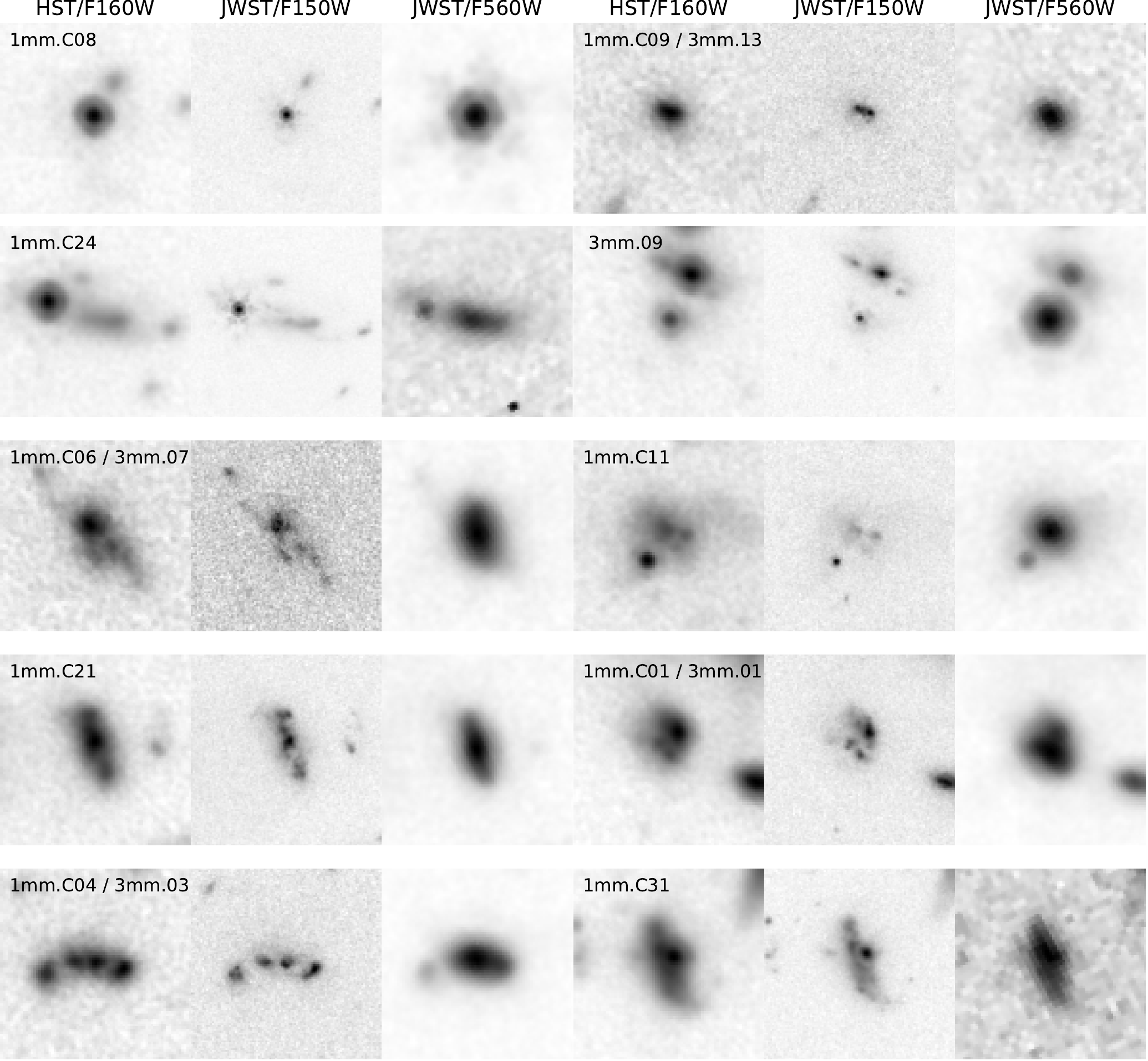}
\caption{$3\arcsec\times3\arcsec$ images of the gas- and dust-mass
  selected galaxies at $z\approx2.5$ at
  $\lambda_{\rm obs} \approx 1.6$\,\micron\ (HST/WFC3 and
  JWST/NIRCam) and $5.6$\,\micron\ (JWST/MIRI) highlight change in
  morphology between the rest-frame UV/optical
  ($\lambda_{\rm rest}\approx 450$\,nm) and rest-frame NIR at
  1.6\,\micron, which is now resolved with MIRI for the first
  time.  The difference in morphology is not due to the point spread
  function, which is very similar between F160W ($\approx0\farcs15$)
  and F560W ($\approx 0\farcs2$)
  \label{fig:sel-cutouts}
}
\end{figure*}

Photometry is performed for all sources in the F560W image in the
MIRI, NIRCam and HST bands with \textsc{The Farmer}
\citep{Weaver2022}, using our F560W PSF model for MIRI (see
\autoref{sec:psf}) and \textsc{WebbPSF} models for NIRCam
\citep{Perrin2012, Perrin2014}.  In brief, we first detect and model
the galaxies in F560W MIRI band, and then perform forced photometry in
the other multi-wavelength bands, allowing the flux to vary, whilst
keeping the structural parameters fixed (details will be presented in
Gillman et al., in prep.).

We compare the F560W fluxes for all sources in the combined MIDIS and
ASPECS field to the Spitzer IRAC/Channel 3 (5.8\micron) fluxes from
CANDELS \citep{Guo2013} in \autoref{fig:miri-spitzer}, masking sources
near the edge of the MIDIS image.  Overall, the Spitzer fluxes agree
well with MIRI at flux densities $\leq 10$\,$\mu$Jy, but show larger
scatter below $\approx1$\,$\mu$Jy, consistent with the 5\,$\sigma$
limit of the Spitzer photometry.  Visual inspection shows that the
strong outliers above the Spitzer limit are largely due to
neighbouring sources that are poorly (or not) deblended in the Spitzer
photometry.  The strongest outlier from ASPECS is 1mm.C22, which is
indeed heavily blended with a similarly-bright neighbouring galaxy
(see \autoref{fig:cutouts}).  Interestingly, towards the bright end
there seems to be a increasing offset between the Spitzer compared to
the MIRI flux (though the number of sources is limited), that has
become more apparent with the updated MIRI F560W flux calibration
released in September
2023.\footnote{\url{https://www.stsci.edu/contents/news/jwst/2023/updates-to-the-miri-imager-flux-calibration-reference-files}}
Potential causes for the offset are discussed in more detail in
\"{O}stlin et al. (in prep.) and are potentially related to the
significant changes in the intrinsic morphology between the HST/F160W
model used to measure the IRAC fluxes and the actual morphology at
F560W micron (see \autoref{sec:rest-frame-morph}) that is used by
\textsc{The Farmer} in this work, as well as deblending issues of
nearby faint galaxies; though the exact nature remains unclear at the
time of writing.  Discrepancies between the Spitzer IRAC and JWST
fluxes at the bright end have also been reported in other works
\citep[e.g.][]{Rieke2023, Yang2023} and at least in-part attributed to
the order-of-magnitude difference in the PSF between IRAC and JWST.

We model the spectral energy distributions using the high-z version of
\textsc{Magphys} \citep{daCunha2008, daCunha2015}.  We join the new
JWST and HST photometry with the total fluxes from existing
longer-wavelength photometry from Spitzer/IRAC 24\,\micron\
\citep{Whitaker2014}, Herschel/PACS 100 and 160\,\micron\
\citep{Elbaz2011} and ALMA at 1\,mm and 3\,mm \citep{Dunlop2017,
  Gonzalez-Lopez2019, Gonzalez-Lopez2020}, including $5\sigma$ upper
limits on the mm flux. The resulting stellar masses are listed in
\autoref{tab:sources}. The total stellar masses are on average very
consistent (within 0.04\,dex) with the pre-JWST determinations using
the same models \citep[cf.][]{Aravena2020, Boogaard2020}, though some
individual sources show larger scatter ($\approx$0.25\,dex) likely due
to differences in the photometry.  We attribute the overall agreement
in the total stellar masses to the relative consistency between the
integrated fluxes from Spitzer and JWST and the large number of
constraints on the shape of the spectral energy distribution across
wavelengths available for the ALMA sources in the HUDF.

\subsection{Structural parameters}
\label{sec:struct-param}

We measure the global structural parameters observed in the MIRI/F560W
filter by modeling the galaxies with a single S\'{e}rsic profile
\citep{Sersic1963} using \textsc{galfit} \citep{Peng2002}.  We create
$200\times200$ pixel cutouts around each source from the
background-subtracted F560W science image and error map at a pixel
scale of $0\farcs06$.  We adopt initial guesses on the magnitude, half
light radius ($R_e$), S\'{e}rsic index ($n$), axis ratio ($b/a$), and
position angle from \textsc{The Farmer} catalog.  We simultaneously
model all sources in the cutout up to 2.5 magnitudes fainter and
within 3\arcsec\ of the target source and mask fainter sources using
the segmentation map.  We slightly tweak these parameters for the
background sources in a few individual cases to improve the overall
fit.  For the PSF, we use our empirical, position-dependent PSF model,
that is described in \autoref{sec:psf}.  To take into account
underestimated random and systematic errors as well as residual
uncertainties in the PSF, we quadratically fold in a minimum relative
uncertainty on the effective radius and S\'{e}rsic index of 5\%,
following \citet[][based on the S/N of the faintest objects in the
sample]{vanderWel2012}.

The best-fit parameters for all ASPECS galaxies are listed in
\autoref{tab:sources}.  We discard the two AGN that are best-fit with
a point-source template from the structural analysis (3mm.09 and
1mm.C08, see \autoref{fig:cutouts}).  From the reference sample of
galaxies in the HUDF (see \autoref{sec:struct-param-z14}), we remove
all galaxies for which \textsc{galfit} returned bad flags or parameter
values at the edge of the parameter space (indicative of bad fits).
These galaxies are typically at the edge of the field and/or affected
by strong gradients in the background.

\section{Results}
\label{sec:results}
\begin{figure*}[t]
  \centering
  \includegraphics[width=\textwidth]{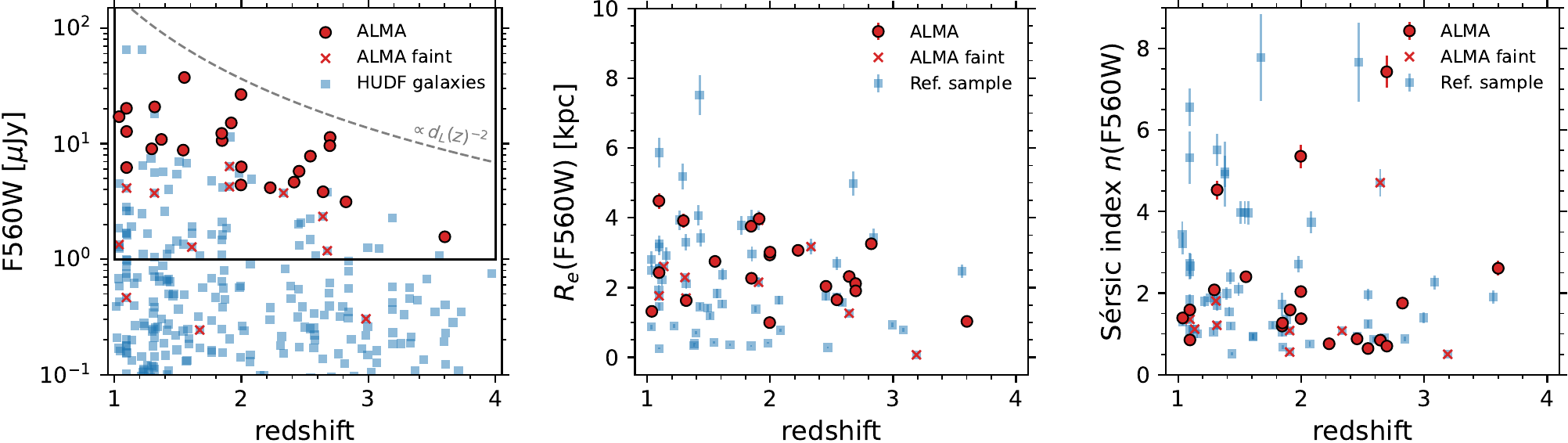}
\caption{ALMA galaxies in context of the galaxy population in the Hubble Ultra Deep Field covered by both MIDIS and ASPECS.  The panels show the MIRI/F560W flux density (\textit{left}), effective radius (\textit{center}) and S\'{e}rsic index (\textit{right}) as a function of redshift.  The black box in the left panel denotes the HUDF reference sample of galaxies with a flux density in F560W $\geq 1.0$\,$\mu$Jy, shown in the other panels.
  \label{fig:refsample-z-vs-mag}}
\end{figure*}

\subsection{Rest-frame Near-Infrared Morphology at $z\approx2.5$}
\label{sec:rest-frame-morph}
The MIDIS observations at 5.6\,\micron\ resolve the rest-frame
near-infrared light of galaxies at $z\approx2.5$, around
$\lambda_{\rm rest}\approx1.6$\,\micron, for the first time.  These
wavelengths trace the bulk of the (older) stellar light, that remained
inaccessible with HST, and are less affected by dust attenuation than
observations at shorter wavelengths.

\autoref{fig:cutouts} shows that the observed-frame 5.6\,\micron\
morphology of the galaxies at $z\approx2.5$ is markedly different from
the morphology traced at shorter wavelengths.  To examine the
morphological differences at $z\approx2.5$ in more detail, we compare
the HST/F160W, and now higher-resolution NIRCam/F150W at similar
wavelengths, to the MIRI/F560W observations in
\autoref{fig:sel-cutouts} (more extensive multi-wavelength cutouts of
all sources are shown in \autoref{sec:multi-wavel-cuto}).  We also
show the higher-redshift sources at $z\approx3.7$ in the figure, where
F560W probes $\lambda_{\rm rest}\approx1.2$\,\micron.  The
\textsc{galfit} models and residuals for the galaxies are shown in
\autoref{fig:galfit-residuals} in \autoref{sec:galfit-residuals}.

The MIRI observations reveal that all $z\approx2.5$ sources have a
centrally concentrated and relatively smooth light distribution at
rest-frame 1.6\,\micron.  This is in contrast to the significantly
more substructured or clumpy appearance in HST and NIRCam at
rest-frame 450\,nm.  We stress the smoother morphology is intrinsic
and not primarily due to the MIRI/F560W PSF (with a
$\mathrm{FWHM}\approx0.2\arcsec$).  This can be seen by comparing to
the HST/F160W observations, which trace roughly the same wavelength as
NIRCam/F182M, but have a similar PSF size
($\mathrm{FWHM}\approx0.15\arcsec$) as MIRI/F560W.  The F560W images
are on the other hand not perfectly smooth and (residual) structure
(after subtracting a single S\'{e}rsic component) can be seen in
\autoref{fig:galfit-residuals}.  However, in all cases, the galaxies
show significantly more substructure in the F160W filter than in
F560W.

As an aside, we note that the high-resolution NIRCam imaging reveals
some striking details in the galaxies
(cf. \autoref{fig:full-cutouts}).  This includes rich substructure in
the X-ray AGN 1mm.C01, which suggests merger activity may be
triggering the starburst and AGN activity in this system.  For the
X-ray AGN 3mm.09, there is extended emission in the F182M filter not
present in F210M (\autoref{fig:full-cutouts}).  This is likely
\Oiii+\Hb\ emission in the medium band that could originate from a gas
in/outflow or gas that is ionised by the AGN at larger distances.
Notable are also the pronounced obscuring dust structures (lanes)
visible in several galaxies across the entire redshift range of our
sample (cf. \autoref{fig:cutouts}).

\subsection{Structural parameters at $z=1$--$4$ with MIRI and HST}
\label{sec:struct-param-z14}

\begin{figure*}[t]
  \centering
  \includegraphics[width=0.9\textwidth]{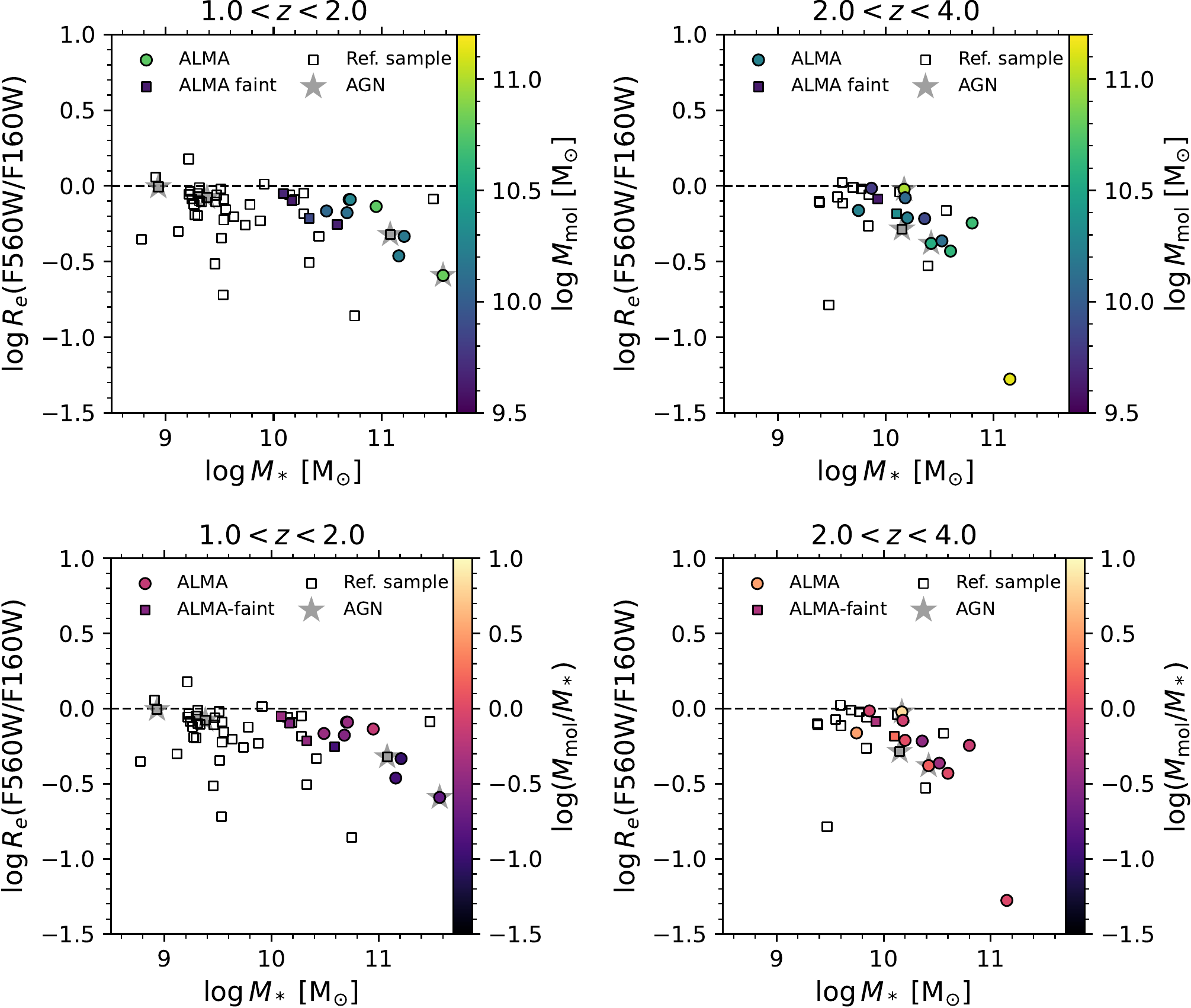}
  \caption{Ratio of the sizes (effective radii) at 5.6\micron\ and
    1.6\micron\ versus stellar mass, color coded by cold gas mass
    (derived from the cold dust emission at 1.2\,mm; see
    \autoref{sec:sample}) (top) and cold gas-to-stellar mass ratio
    (bottom) for the ALMA galaxies in the context of the HUDF
    reference sample (see \autoref{fig:refsample-z-vs-mag}).  The left
    and right panels show the redshift ranges $1.0<z<2.0$ and
    $2.0<z<4.0$ respectively.  The size ratio anti-correlates clearly
    with stellar mass.  It also shows a potential (anti-)correlation
    with gas-to-stellar mass ratio (gas mass), see
    \autoref{fig:size-mmol-mstar}. \label{fig:size-mstar-mmol}}
\end{figure*}

\begin{figure*}[t]
  \centering
  \includegraphics[width=0.9\textwidth]{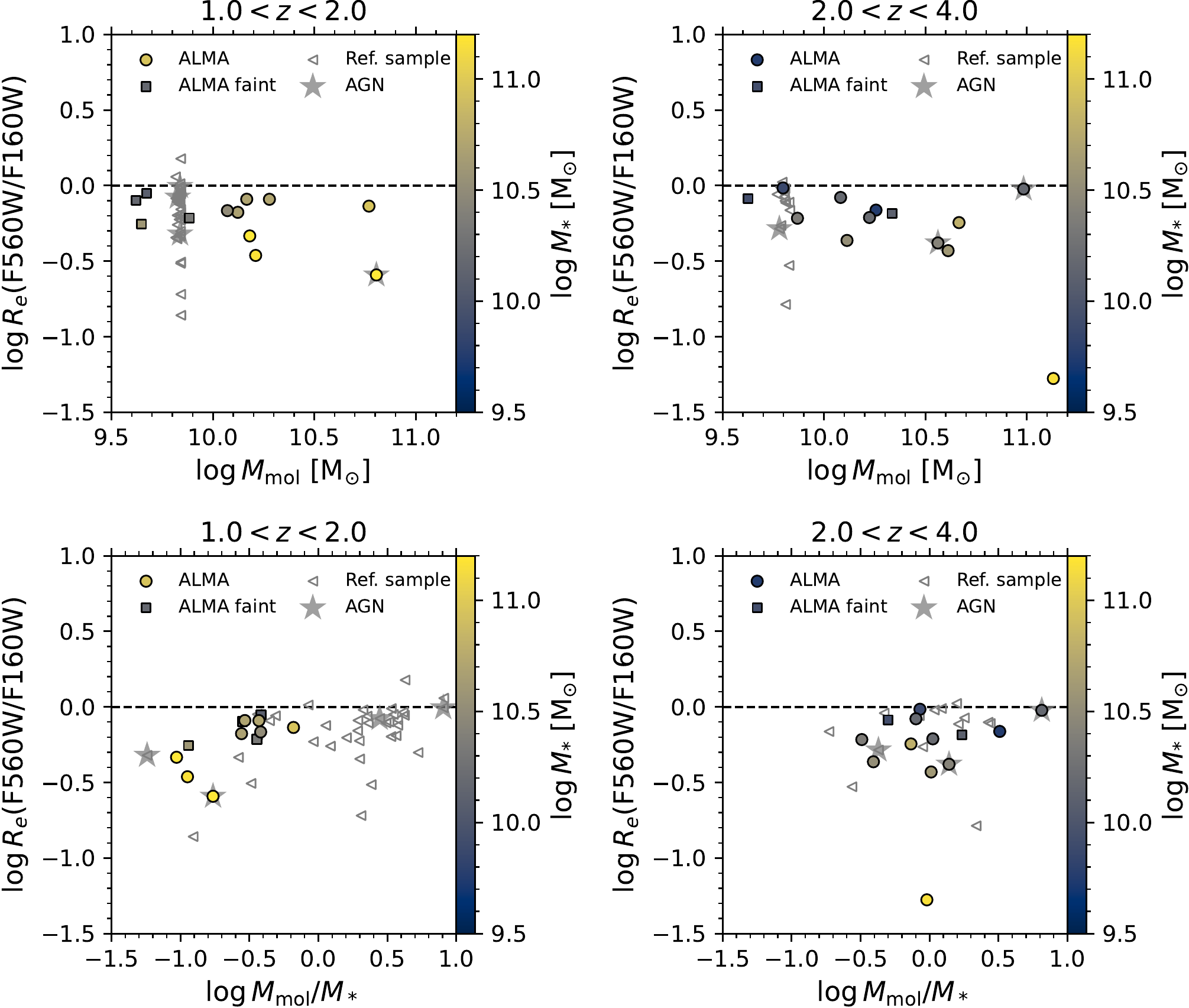}
  \caption{Ratio of the sizes (effective radii) at 5.6\micron\ and
    1.6\micron\ versus cold gas mass (derived from the cold dust
    emission at 1.2\,mm; see \autoref{sec:sample}) (top) and
    gas-to-stellar mass ratio (bottom), color coded by stellar mass
    for the ALMA galaxies.  Upper limits on the gas mass and
    gas-to-stellar mass ratio are shown for the HUDF reference sample.
    The left and right panels show the redshift ranges $1.0<z<2.0$ and
    $2.0<z<4.0$ respectively.  The gas-to-stellar mass ratio (gas
    mass) (anti-)correlates more with size ratio, connected to the
    correlation with stellar mass (cf.
    \autoref{fig:size-mstar-mmol}). \label{fig:size-mmol-mstar}}
\end{figure*}

We analyse the structural parameters of the ALMA sources in the
context of a flux-limited reference sample, consisting of all galaxies
at $z=1$--4 in the HUDF galaxy population covered by MIRI and ALMA
that have a flux density $\fnumiri \geq 1.0$\,$\mu$Jy.  The redshifts
for the non-ALMA sources are taken from the MUSE \citep{Bacon2023} or
else 3D-HST \citep{Momcheva2016} catalogs.

We show the F560W fluxes of the dust continuum-selected galaxies in
the context of the HUDF galaxy population in
\autoref{fig:refsample-z-vs-mag}.  The ASPECS galaxies trace the
bright end of the galaxy population in F560W.  In addition to the
flux-limited ALMA sample, we also mark the fainter galaxies in the
MIDIS footprint that were detected at 1.2\,mm based on an optical--FIR
prior \citep{Gonzalez-Lopez2020}.  These galaxies mostly correspond to
galaxies that lie just below the formal blind detection threshold of
ASPECS.

The \textsc{galfit}-derived structural parameters are shown in
\autoref{fig:refsample-z-vs-mag}.  The effective radii of the ALMA
galaxies are between $\Remiri=1$--5\,kpc, slightly decreasing towards
higher redshift on average, as expected from the evolution of the
mass-size relation \citep{vanderWel2014}.  The ALMA galaxies trace a
similar distribution in effective radius as the other galaxies from
the HUDF reference sample.  The S\'{e}rsic indices are shown in the
rightmost panel of \autoref{fig:refsample-z-vs-mag}.  Between $2<z<3$,
where MIRI traces rest-frame
$\lambda_{\rm rest} \approx 1.6$\,\micron, most galaxies have
$n\approx1$ indicating an exponential profile that is typical for
star-forming disks, while a few show higher, $n\geq2$, S\'{e}rsic
indices.  The higher-redshift galaxy at $z=3.6$
($\lambda_{\rm rest} \approx 1.2$\,\micron) has $n\approx2.6$.  Below
$z=2$, the ALMA galaxies show a range of S\'{e}rsic indices, mostly
between 1 and 3.  Overall, the distribution of S\'{e}rsic indices
follows the HUDF reference sample, with potentially a slight trend
towards lower S\'{e}rsic indices.

Next, we compare the structural parameters measured at
$\lambda_{\rm obs} = 5.6$\,\micron\ by MIRI to the measurements at
$\lambda_{\rm obs} = 1.6$\,\micron.  As measuring the structural
parameters requires accurate knowledge of the PSF,
we do not measure the structural parameters from the NIRCam imaging,
but instead refer to the established measurements from HST/F160W by
\cite{vanderWel2012}.  This excludes 1mm.C12 (AGN) and 1mm.C22 as they
are not properly deblended in the HST
catalog.

The average effective radius at 5.6\,\micron\ is smaller than at
1.6\,\micron.  For the ASPECS sources, \Remiri\ is on average about
35\% and 30\% smaller than \Rehst\ at $1<z<2$ and $2<z<4$,
respectively.  For the full reference sample at
$\fnumiri \geq 1$\,$\mu$Jy, \Remiri\ is about 20\% smaller than
\Rehst\ at both $1<z<2$ and $2<z<4$.  We show the ratio of effective
radii as a function of stellar mass $1<z<2$ and $2<z<4$ in
\autoref{fig:size-mstar-mmol}.  We find that the ratio decreases as a
function of stellar mass, such that more massive galaxies have
relatively more compact light profiles at longer wavelengths.  A
similar trend has been shown by \cite{Suess2022} comparing
shorter-wavelength NIRCam 4.4 and 1.5\,\micron\ observations for a
larger sample of galaxies out to $z=2.5$ (see also \citealt{Chen2022,
  Gillman2023}).  The S\'{e}rsic indices are, on average, roughly a
factor $\approx1.5$ larger in F560W than in F160W, though with a
significant scatter and no clear trend with stellar mass or redshift.

We also investigate correlations between the JWST/HST size ratio and
quantities related to the cold gas mass (or, equivalently, the cold
dust mass, see \autoref{sec:sample}) in \autoref{fig:size-mmol-mstar}.
We find that the size-ratio weakly anti-correlates with cold gas (or
cold dust) mass, but more strongly correlates with the cold
gas-to-stellar mass ratio, with a slope that is close to the inverse
of the trend found with stellar mass.

\section{Discussion} \label{sec:discussion}
\subsection{HST vs.\ JWST structure and morphology}

The MIRI observations imply that the mass-weighted stellar structure
($\lambda_{\rm rest} \approx 1.6$\,\micron) of these galaxies is
significantly smoother compared to the light-weighted structure in the
rest-frame optical frame observed with HST and NIRCam at
$\lambda_{\rm rest} \approx 450$\,nm.  The more structured and clumpy
rest-optical morphology can be due to a combination of different
factors.  If the underlying true distribution of young stars and star
formation (traced by HST) was as smooth as seen in the stellar mass
(traced by JWST/MIRI), the structure at 450\,nm could be due to a
patchy distribution of attenuating dust, leading to the more clumpy
morphology.  On the other hand, star formation is expected to occur in
spatially-separated clumps, i.e., the observed HST morphology may
reflect the actual underlying distribution of young stars.  Matched
high-resolution ALMA imaging of the dust distribution will be able to
measure the relative impact of the two scenarios.

\subsection{HST vs.\ JWST sizes and color gradients}
Most galaxies in our sample at $z\approx2.5$ show large
$\Remiri \approx 1$--5\,kpc, exponential (disk-like) structures
($n\approx1$).  The fact that the ASPECS galaxies do not clearly stand
out in their structural parameters compared to the other sources in
the HUDF reference sample (\autoref{fig:refsample-z-vs-mag}), implies
that at the depth reached by ASPECS over a relatively small field, the
galaxies mostly trace the massive end of the typical galaxy population
at these redshifts \citep[cf.][]{Boogaard2019, Gonzalez-Lopez2020}.

The inferred rest-frame near-infrared sizes are smaller compared to
those observed in the rest-frame optical, implying a negative color
gradient (i.e., the galaxies have redder centers).  A similar trend
has also been observed in other galaxy populations
\citep[e.g.,][]{Suess2022, Chen2022, Gillman2023, Magnelli2023},
including local face-on spiral galaxies \citep[e.g.,][]{Casasola2017},
and in studies of stellar mass maps
\citep[e.g.,][]{Wuyts2012,vanderWel2023}, and can be driven by
differences in the properties of the stars, such as their age, or by
dust extinction.  The results can thus be attributed to the presence
or formation of an older, more centrally concentrated stellar
population (such as a bulge) and/or stronger dust extinction in the
center (potentially linked to compact central star formation), where
both lead to a `flatter' distribution in the rest-frame optical bands.
The trend with stellar mass can be linked to the same effects becoming
stronger in more massive star-forming galaxies.  Indeed, massive
galaxies are known to show overall stronger extinction and a larger
fraction of obscured star formation \citep[e.g.,][]{Garn2010,
  Whitaker2017} and have stronger extinction towards their centers
\citep[e.g.,][]{Nelson2016a, Matharu2023}.

While the rest-frame 1.6\,\micron\ emission is primarily sensitive to
the old stellar light, it may also contain a contribution from an AGN,
which can make the profiles look more centrally concentrated
\citep[e.g.,][]{Prieto2010}.  While the number of AGN in the sample is
limited, we do not see a clear difference in
\autoref{fig:size-mstar-mmol} between the ALMA galaxies that are
identified as AGN and those that are not (note this already excludes
the two AGN best-fit with a point-source model).  In the same vein,
the MIRI emission may also trace nebular continuum and/or high
equivalent width emission lines in case of very vigorous young star
formation, which can have a complicated impact on the morphology
\citep[e.g.,][]{Papaderos2023}.  This would most strongly impact
galaxies with a low stellar mass and/or high specific SFR.  Given the
comparatively smooth light distribution over large scales, this
however does not appear to have a major impact on the MIRI morphology
of the relatively massive gas rich galaxies studied here.

\subsection{Trends with  total gas/dust mass?}
It would be interesting to assess whether the color and size trends
above are correlated with any other galaxy property.  As dust can be
responsible for some of the observed trends, we can check for the
influence of the {\em total} dust (or gas) mass, even in the absence
of resolved dust imaging (see \autoref{fig:size-mstar-mmol}).
However, both the gas mass and the gas-to-stellar-mass fraction are
known to independently correlate and anti-correlate with stellar mass,
respectively \citep[e.g.,][]{Tacconi2018, Aravena2019}, and larger
samples are needed to distinguish potential trends.  Again, we would
expect a relation with the resolved dust properties.  The sizes of the
dust in these massive star-forming galaxies are often significantly
smaller than the rest-optical sizes \citep[][]{Tadaki2020}, though
typically not extremely compact as seen in submillimeter galaxies
\cite[e.g.][]{Gullberg2019}.  Based on the average sub-mm/optical size
ratios, they may even be more compact than the stellar sizes now
measured with MIRI.  This centrally concentrated dust emission is
often linked to a compact starburst, which may be responsible for
building up bulges \citep[e.g.][]{Nelson2019, Tadaki2020}.  Though
note the dust may also appear more compact due to dust temperature
gradients, as also shown in simulations \citep[e.g.,][]{Cochrane2019,
  Popping2022}.  Matched high-resolution ALMA imaging, especially in
multiple bands, would help to differentiate between the presence (or
formation) of a bulge and/or stronger extinction in the center,
especially in combination with studies of the resolved color gradients
and spectral energy distributions now possible with JWST
\citep[e.g.][]{Miller2022, Perez-Gonzalez2023a}.

\section{Summary and Conclusions} \label{sec:conclusions}

We present JWST/MIRI F560W observations of the stellar structure of
gas- and dust-rich galaxies in the Hubble Ultra Deep Field (HUDF) at
rest-wavelengths of $\lambda>1$\,\micron\ using the MIRI Deep Imaging
Survey (MIDIS).

We select a complete, 1.2\,mm continuum flux-limited sample of 35
galaxies from the ALMA Spectroscopic Survey (ASPECS)---encompassing
all sources from shallower ALMA 1\,mm continuum imaging in the
HUDF---of which 28 lie within the 4.7\,arcmin$^{2}$ footprint of
MIDIS, at $z=0.5$--3.7.  Using JWST slitless spectroscopy, we
determine spectroscopic redshifts for all of the few galaxies in the
ASPECS sample that were still missing spectroscopic confirmation, in
particular in the $2<z<3$ range.

We find reasonable agreement between the MIRI/F560W flux densities and
those previously determined from (deblended) Spitzer IRAC
5.8\,\micron\ observations for the ALMA sources, though with a
potential systematic offset at the bright end.  Subsequently, we
revisit the stellar masses by modeling the spectral energy
distributions with \textsc{Magphys} models, finding good agreement
with previous determinations.  The reinforcement of the stellar mass
estimates for the ALMA galaxies at $z=1-4$ implies that there are no
fundamental changes to the previously reported gas and dust-mass
fractions (for these bright and relatively massive systems), nor to
how the properties of this population of gas and dust-rich galaxies
evolve with stellar mass \citep[][]{Aravena2019, Aravena2020,
  Boogaard2019, Gonzalez-Lopez2020}.

We find that the rest-frame near-infrared light distribution at
$\lambda_{\rm rest} \approx 1.6$\,\micron\ of the ALMA galaxies at
$z\approx2.5$---that can now be resolved with MIRI---is intrinsically
more smooth and centrally concentrated compared to the more
substructured or clumpy appearance in the rest-frame UV/optical at
$\lambda_{\rm obs} \approx 450$\,nm as probed by HST/F160W. This is
not a resolution effect as both observations have a similar PSF.  We
build a custom, position-dependent, empirical PSF model for the MIDIS
observations (described in \autoref{sec:psf}) and use it to perform a
structural analysis using \textsc{galfit}.  We find the galaxies at
$z\approx2.5$ have effective radii of $\Remiri=1$--5\,kpc and
S\'{e}rsic indices mostly close to $n=1$, consistent with an
exponential (disk-like) profile, up to $n\approx5$ (excluding AGN).
We find average size ratios between JWST and HST of
$\Remiri/\Rehst\approx0.75$ and 0.7 at $z\approx1.5$ and
$z\approx 2.5$ respectively, that decrease with stellar mass.  Overall
the mass-weighted stellar structure of the ALMA-selected galaxies is
indistinguishable from a reference sample of other galaxies in the
HUDF.

The results imply that a smoother stellar structure is already in place
in gas-rich, star-forming galaxies at cosmic noon and their clumpy
rest-frame optical is likely caused by a combination of intrinsically
clumpy star formation and/or patchy dust extinction.  The difference
in the mass-weighted radial structure of the galaxies now traced by
MIRI, compared to earlier observations with HST, can be explained both
by the presence of stronger dust extinction; in the center,
potentially linked to compact central star formation, and/or presence
or formation of older central stellar populations, such as a bulge.
Future matched-resolution, spatially resolved ALMA observations will
be key to measure the actual extent (size) of the dust and its
resolved structure and column densities on the same scale as MIRI, and
link the dust properties to the resolved optical/near-IR morphology
that we can now for the first time characterize with JWST.

\facilities{HST, JWST, ALMA}

\software{\textsc{topcat} \citep{taylor2005}, \textsc{gnuastro}
  \citep{akhlaghi2015}, \textsc{ipython} \citep{perez2007},
  \textsc{numpy} \citep{numpy2020}, \textsc{scipy} \citep{scipy2020},
  \textsc{matplotlib} \citep{hunter2007}, \textsc{astropy}
  \citep{TheAstropyCollaboration2022}, \textsc{grizli} \citep{Grizli2023},
  \textsc{photutils} \citep{photutils160}, \textsc{interferopy}
  \citep{interferopy}.}

\begin{acknowledgments}
  \footnotesize \textit{Acknowledgements.}  We want to thank the
  referee for a constructive report that helped improve the paper.  LB
  wants to thank Jeroen Bouwman and James Davies for insightful
  discussions on the MIRI PSF and JWST pipeline.  LB wants to thank
  Pascal Oesch for the FRESCO spectra and Fengwu Sun for providing
  feedback on the redshift estimates.  LB, FW and SB acknowledge
  support by ERC AdG grant 740246 (Cosmic-Gas).  SG acknowledges
  financial support from the Villum Young Investigator grant 37440 and
  13160 and the Cosmic Dawn Center (DAWN), funded by the Danish
  National Research Foundation (DNRF) under grant No. 140.  LC, JAM,
  ACG acknowledge support by grant PIB2021-127718NB-100 from the
  Spanish Ministry of Science and Innovation/State Agency of Research
  MCIN/AEI/10.13039/501100011033.  KIC, EI and PR acknowledge funding
  from the Netherlands Research School for Astronomy (NOVA).  KIC
  acknowledges funding from the Dutch Research Council (NWO) through
  the award of the Vici Grant VI.C.212.036.  TRG and IJ acknowledge
  support from the Carlsberg Foundation (grant no CF20-0534).  PGP-G
  and LC acknowledge support from grant PID2022-139567NB-I00 funded by
  the Spanish Ministry of Science and Innovation/State Agency of
  Research MCIN/AEI/10.13039/501100011033.  AAH acknowledges support
  from grant PID2021-124665NB-I00 funded by
  MCIN/AEI/10.13039/501100011033 and by ERDF A way of making Europe.
  AB acknowledges support from the Swedish National Space
  Administration (SNSA).  LC acknowledges financial support from
  Comunidad de Madrid under Atracci\'on de Talento grant
  2018-T2/TIC-11612.  AE and FP acknowledge support through the German
  Space Agency DLR 50OS1501 and DLR 50OS2001 from 2015 to 2023.  JH
  and DL were supported by a VILLUM FONDEN Investigator grant (project
  number 16599).  RAM acknowledges support from the Swiss National
  Science Foundation (SNSF) through project grant 200020\_207349.  GW,
  DD, JPP, TT acknowledge financial support from the UK Science and
  Technology Facilities Council, and the UK Space Agency.

  The work presented is the effort of the entire MIRI team and the
  enthusiasm within the MIRI partnership is a significant factor in
  its success. MIRI draws on the scientific and technical expertise of
  the following organisations: Ames Research Center, USA; Airbus
  Defence and Space, UK; CEA-Irfu, Saclay, France; Centre Spatial de
  Liége, Belgium; Consejo Superior de Investigaciones Científicas,
  Spain; Carl Zeiss Optronics, Germany; Chalmers University of
  Technology, Sweden; Danish Space Research Institute, Denmark; Dublin
  Institute for Advanced Studies, Ireland; European Space Agency,
  Netherlands; ETCA, Belgium; ETH Zurich, Switzerland; Goddard Space
  Flight Center, USA; Institute d'Astrophysique Spatiale, France;
  Instituto Nacional de Técnica Aeroespacial, Spain; Institute for
  Astronomy, Edinburgh, UK; Jet Propulsion Laboratory, USA;
  Laboratoire d'Astrophysique de Marseille (LAM), France; Leiden
  University, Netherlands; Lockheed Advanced Technology Center (USA);
  NOVA Opt-IR group at Dwingeloo, Netherlands; Northrop Grumman, USA;
  Max-Planck Institut für Astronomie (MPIA), Heidelberg, Germany;
  Laboratoire d’Etudes Spatiales et d'Instrumentation en Astrophysique
  (LESIA), France; Paul Scherrer Institut, Switzerland; Raytheon
  Vision Systems, USA; RUAG Aerospace, Switzerland; Rutherford
  Appleton Laboratory (RAL Space), UK; Space Telescope Science
  Institute, USA; Toegepast- Natuurwetenschappelijk Onderzoek
  (TNO-TPD), Netherlands; UK Astronomy Technology Centre, UK;
  University College London, UK; University of Amsterdam, Netherlands;
  University of Arizona, USA; University of Cardiff, UK; University of
  Cologne, Germany; University of Ghent; University of Groningen,
  Netherlands; University of Leicester, UK; University of Leuven,
  Belgium; University of Stockholm, Sweden; Utah State University,
  USA. A portion of this work was carried out at the Jet Propulsion
  Laboratory, California Institute of Technology, under a contract
  with the National Aeronautics and Space Administration. We would
  like to thank the following National and International Funding
  Agencies for their support of the MIRI development: NASA; ESA;
  Belgian Science Policy Office; Centre Nationale D'Etudes Spatiales
  (CNES); Danish National Space Centre; Deutsches Zentrum fur Luft-und
  Raumfahrt (DLR); Enterprise Ireland; Ministerio De Econom\'ia y
  Competitividad; Netherlands Research School for Astronomy (NOVA);
  Netherlands Organisation for Scientific Research (NWO); Science and
  Technology Facilities Council; Swiss Space Office; Swedish National
  Space Board; UK Space Agency.

  This work is based on observations made with the NASA/ESA/CSA James
  Webb Space Telescope. Some data were obtained from the Mikulski
  Archive for Space Telescopes at the Space Telescope Science
  Institute, which is operated by the Association of Universities for
  Research in Astronomy, Inc., under NASA contract NAS 5-03127 for
  \textit{JWST}; and from the
  \href{https://jwst.esac.esa.int/archive/}{European \textit{JWST}
    archive (e\textit{JWST})} operated by the ESDC.  The observations
  can be accessed via \url{http://dx.doi.org/10.17909/gn5v-f189}.

\end{acknowledgments}\vspace{-1cm}

\input{sources-final.tex}

\bibliography{library}{}
\bibliographystyle{aasjournal}

\appendix

\section{MIRI/F560W Point Spread Function}
\label{sec:psf}
The MIRI point spread function (PSF) at 5.6\,\micron\ is affected by a
number of effects, including the non-linearity of the detector, the
brighter-fatter effect, and internal diffraction that occurs inside
the detector below approximately 10\,\micron\
\citep[][cf. \citealt{Wright2023}]{Gaspar2021, Agryiou2023}, which are
are not fully captured by the current \textsc{WebbPSF}
models.\footnote{\url{https://github.com/spacetelescope/webbpsf/releases/tag/v1.2.1}}
While the first two are mainly relevant in the case of very bright
sources---of which there are none in the MIDIS field---the latter
effect can scatter photons to large distances from the core of the PSF
and is very relevant for observations at 5.6\,\micron.  The net result
of the internal diffraction is a broadening of the core of the PSF and
a cruciform artefact out to large radii.  The shape of the cruciform
artefact is dependent on the angle of incidence and varies with
position across the detector, with the spikes effectively bending
inwards towards the center of the detector.

\begin{figure*}[t]
  \centering
  \includegraphics[width=0.95\textwidth]{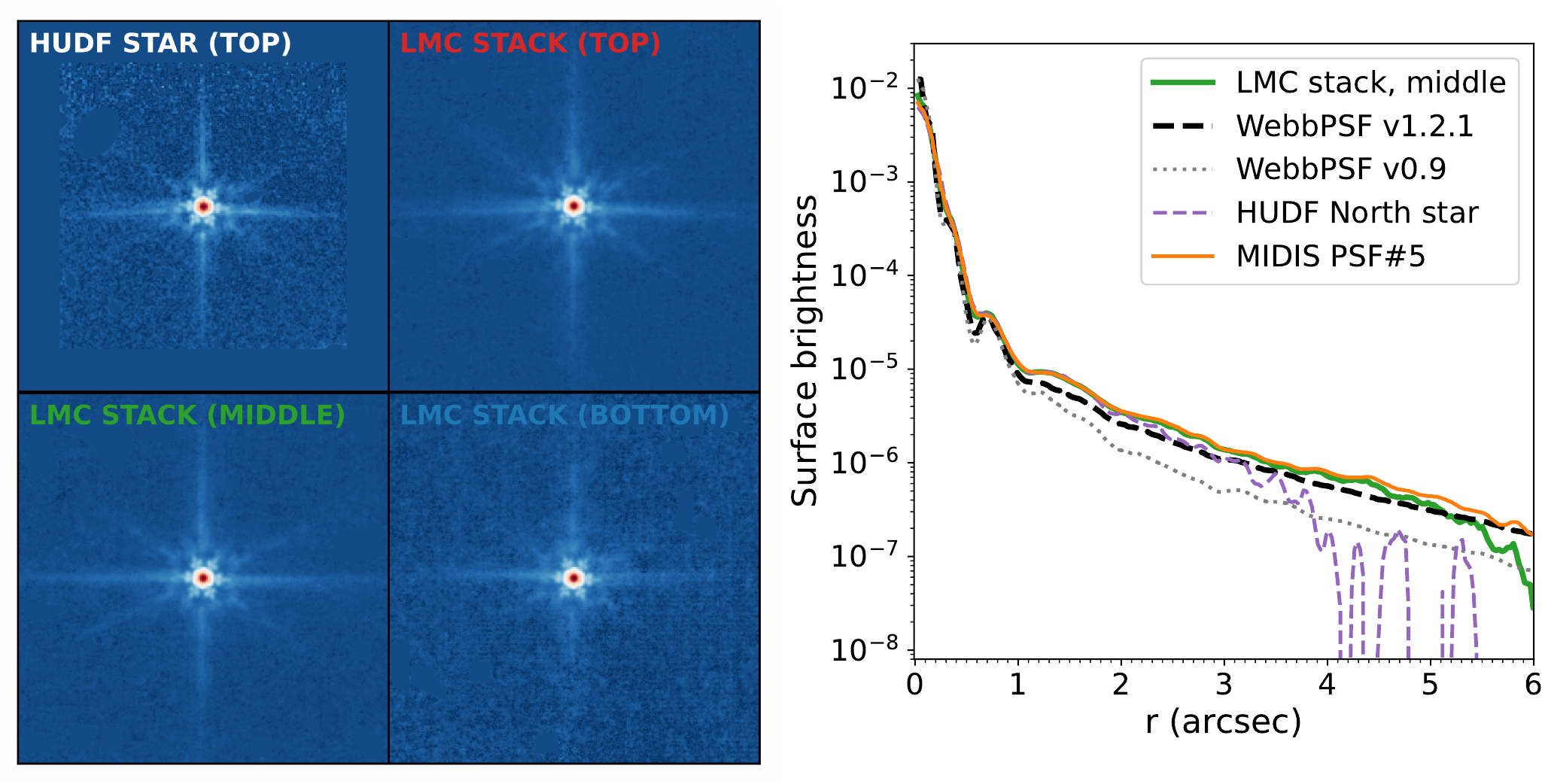}
  \caption{The MIRI/F560W PSF.  \textbf{Left:} The top left panel
    shows a single star in the north of the MIDIS field.  The other
    panels show the stacked PSF from the LMC commissioning data for
    the top-, middle- and bottom-third of the MIRI field
    ($16\farcs5\times16\farcs5$ cutouts with logarithmic
    scaling). \textbf{Right:} Radial profiles of the different PSFs,
    compared to the \textsc{WebbPSF} models (version v0.9 and v1.2.1;
    where the later includes detector effects), on the same pixelscale
    and normalized to have the same total flux within a $1\farcs1$
    aperture.  The LMC-based PSF is smoother and has more power in the
    wings than both the \textsc{WebbPSF} models.  The PSF from the
    star in the north of the field and the MIDIS PSF model (\#5, see
    \autoref{fig:psf-insert}) are in good agreement.  The differences
    in core width are likely due to the (re)sampling of the
    intrinsically undersampled PSFs, compared to the perfectly
    (over)sampled \textsc{WebbPSF} model.  The PSF from the star
    reaches low S/N at large radii (beyond
    $3\farcs5$). \label{fig:psf-model}}
\end{figure*}

\begin{figure*}[t]
  \centering
  \includegraphics[width=0.95\textwidth]{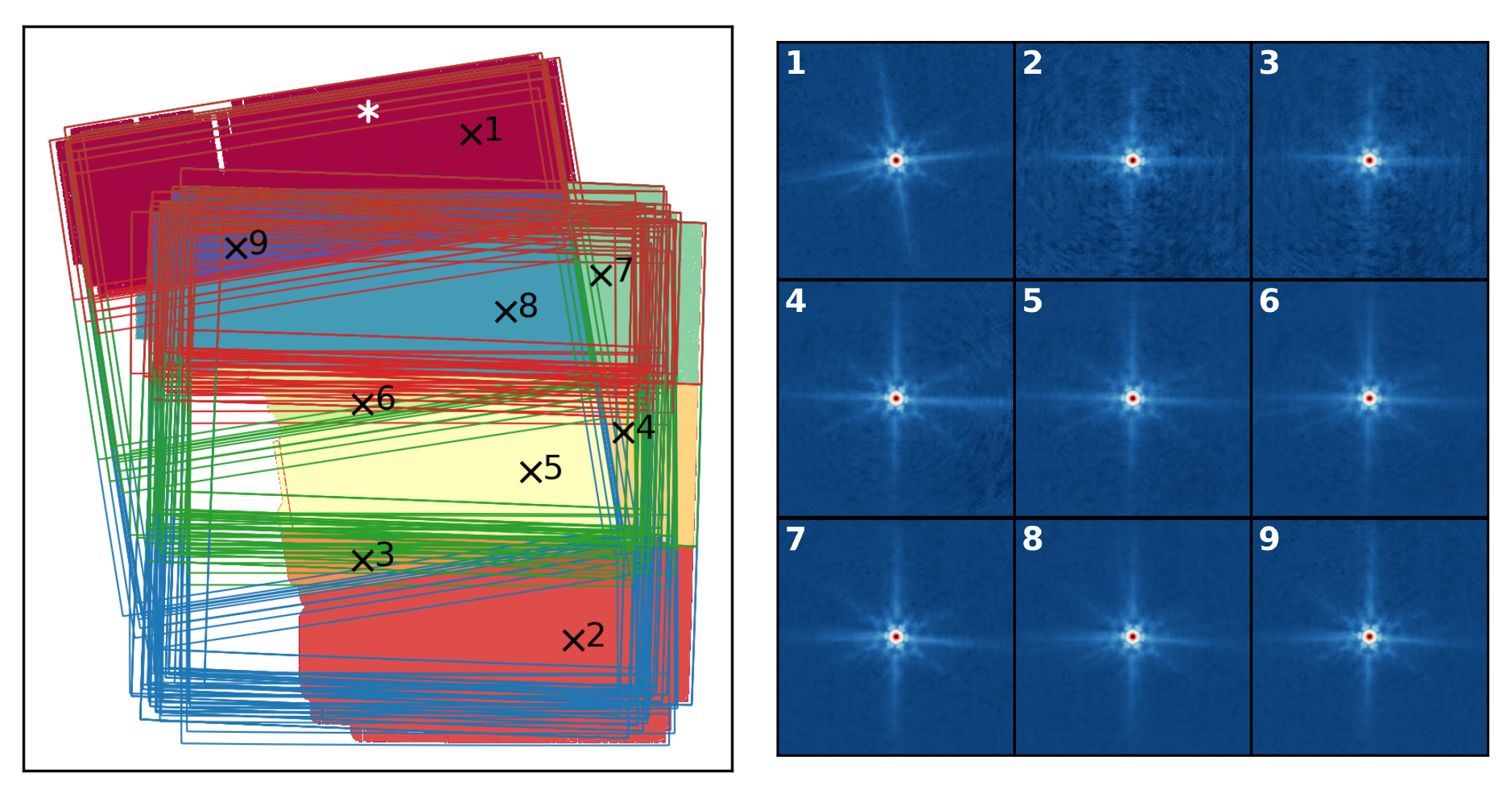}
  \caption{Model of the MIRI/F560 PSF variation over the MIDIS field.
    For each of the 96 exposures, the red, green and blue areas in the
    left panel indicate where the three different stacked PSFs are
    applicable and the white star marks the location of the star in
    the image (see \autoref{fig:psf-model}).  The background color map
    shows the 9 unique areas where the pipeline-processed PSF, which
    is created by inserting the different PSF model in each exposure
    at the marked coordinates, is effective. The right panels show a
    $16\farcs5\times16\farcs5$ cutout of the pipeline-processed PSFs
    at the 9 locations.
    \label{fig:psf-insert}}
\end{figure*}

For these reasons, we construct a varying PSF model for the MIDIS
field making use of empirical PSFs at different positions on the MIRI
detector.  The MIDIS image itself contains only two bright stars, of
which one is in a highly crowded region.  The other bright star is
shown in \autoref{fig:psf-model}, where we have masked background
sources.  This star is in the very north of the field and only in 9\% of all exposures, close to
the top of the detector (see \autoref{fig:psf-insert}), and the bending
of the cruciform artifact can be clearly seen for this star.  Hence,
we cannot use this single star as an empirical PSF for the entire
field.

Instead, we create the empirical PSFs by stacking stars in the Large
Magellanic Cloud (LMC) images from the commissioning program ID 1473
and two calibration programs (IDs: 1040 and 1024).  The F560W frames
in each program are reduced and resampled to a pixel scale of 55 mas
(i.e., oversampled by a factor of 2) using the standard JWST pipeline.
Given that we want to stack many different stars, the astrometry needs
to be thoroughly checked in the image, but since these are well known
regions of the sky, we found the JWST pipeline astrometry to be
sufficient.  We then select a total of 64 stars in the stacked frames.
In this selection we avoid saturated stars and stars in very crowded
regions, i.e., stars which have other sources within $\sim 1\arcsec$.
A stamp of size $\sim16$ arcsec is cut out for each star and a local
background is subtracted.  Faint sources are masked in each stamp
using standard source detection techniques
(\texttt{photutils.detection}).  The images are resampled keeping the
native detector orientation, which enables us to separately stack
three bins of stars, situated in upper, middle, and lower detector
positions.  By doing so we take into account the strongest PSF
variation in the field that occurs over the longest axis of the
detector.  The total number of stars in these bins is 31, 21, and 12,
respectively.  We scale the star stamps to have the same flux within a
radius of 2 pixels (0.11\arcsec, measured using \texttt{photutils})
and stack the stars in each detector position bin by using median
stacking (dropping the 2 brightest and faintest data points).  The
resulting three empirical PSFs still have a few defects in the
outskirts of the PSF image which are manually masked (flux set to
zero).  We inspect signal-to-noise maps (using the standard deviation
of the stacked star stamps as noise), to make sure that the final PSF
model is valid.  There is an insufficient number of stars for a more
detailed spatially varying model using this technique, but we note
that with additional calibration data, and by stacking stars from
individual data frames (before resampling/stacking the full images),
PSF models with higher sub-sampling and similar (or even greater)
depth can be derived, though this is beyond the scope of the present
work. The resulting stacked PSFs and their radial profiles are shown
in \autoref{fig:psf-model}.  These clearly show the up- and downward
bending of the cruciform artifact along the long axis of the detector
and the significantly broader core compared to the \textsc{WebbPSF}
model.  The LMC-based PSF model is available at
\url{https://github.com/jensmelinder/miripsfs}.

We use the empirical PSFs to create a field varying PSF for the MIDIS
field.  Based on the relative orientation of each of the MIDIS
exposures, we determine 9 areas in the MIDIS field between which the
PSF varies most strongly and mark a WCS coordinate ($\alpha, \delta$)
in each area, as shown in \autoref{fig:psf-insert}.  In each of these
areas we insert the appropriate empirical PSF, based on the
corresponding image coordinates $(x,y$), into each of the 96
individual exposures (i.e., WCS-aligned \texttt{cal} files). We then
run the Stage 3 Imaging Processing twice, once with the inserted stars
and once without, and recover the varying PSF model by taking the
difference.  To ensure the pipeline produces identical results in the
different runs, we turn off all steps except the \texttt{resample}
step.  This results in 9 effective PSFs, shown in
\autoref{fig:psf-insert}, that model the largest PSF variations across
the MIDIS field.  The MIDIS PSF model is available at
\url{https://github.com/lboogaard/midis_psf}.  The radial profile of
the MIDIS PSF is very close to the profile of the only sufficiently
bright and isolated real star in the MIDIS field (HUDF North star).
The star as well as the (LMC-based) input PSF model and resulting
MIDIS PSF are smoother in the core region compared to the WebbPSF
models, which is likely a result of (under)sampling compared to the
perfectly (over)sampled WebbPSF models.

\section{FRESCO spectra}
\label{sec:fresco-spectra}
\begin{figure*}[t]
  \includegraphics[width=\textwidth]{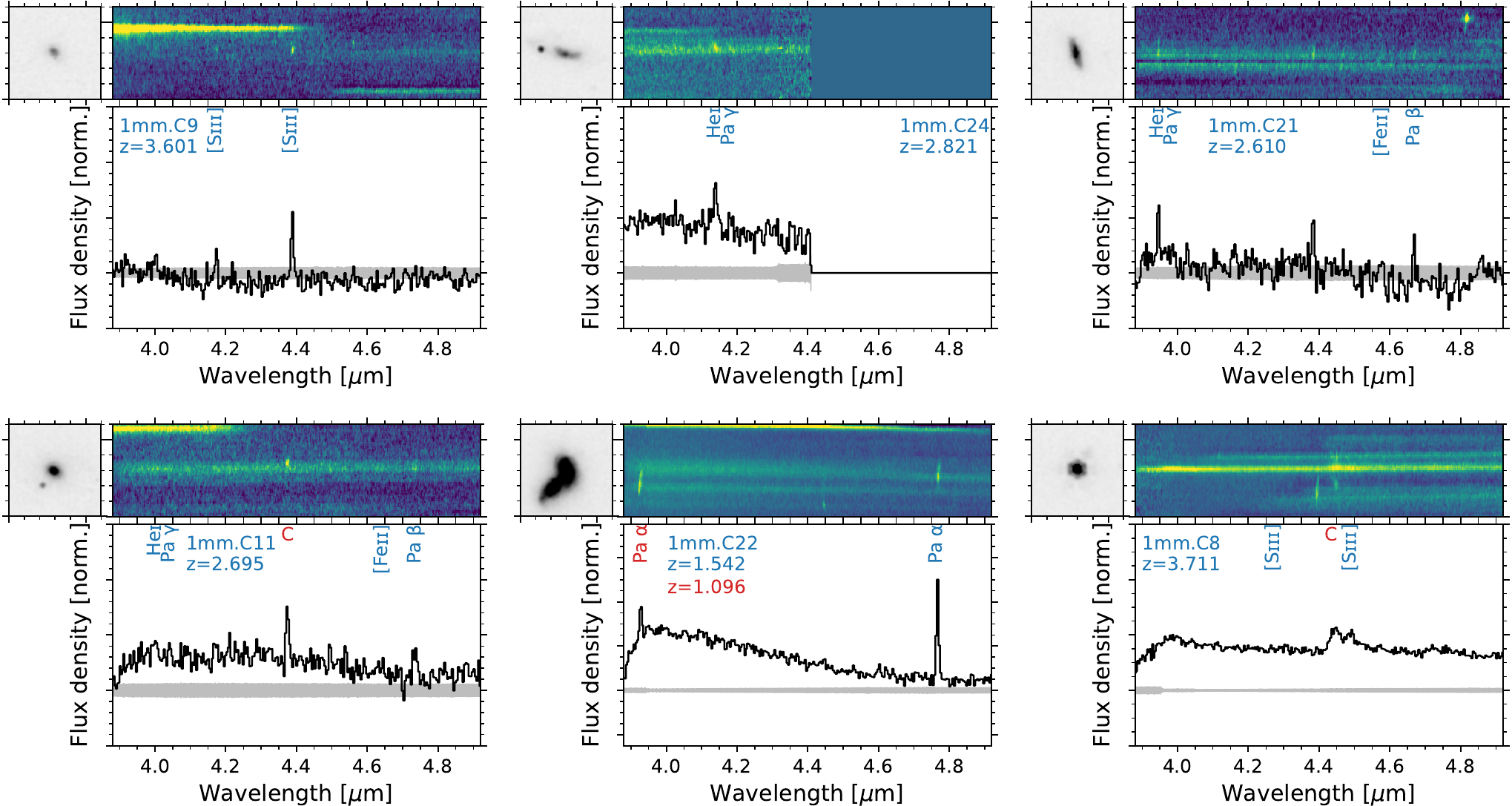}
  \caption{NIRCam slitless spectra for selected sources from FRESCO.
    The top panels show the F444W cutout and 2D spectrum, the bottom
    panel the binned 1D extracted spectrum (the gray shading shows the
    noise).  Lines used to identify the redshift of the ASPECS sources
    are shown in blue (not all detected).  Other (blended) redshifts
    or contamination (C) from neighbouring sources are indicated in
    red.\label{fig:spectra}}
\end{figure*}

We show the FRESCO spectra of the sources discussed in
\autoref{sec:redshifts} in \autoref{fig:spectra}.

\section{Multi-wavelength Cutouts}
\label{sec:multi-wavel-cuto}
We show $4\arcsec\times4\arcsec$ cutouts ($8\arcsec\times8\arcsec$
cutouts for 1mm.C28, 1mm.C30, 1mm.C32 and 1mm.C33 at $z<1$) in all
filters discussed in \autoref{sec:observations} in
\autoref{fig:full-cutouts}.

\begin{figure*}[t]
  \centering
\includegraphics[width=\textwidth]{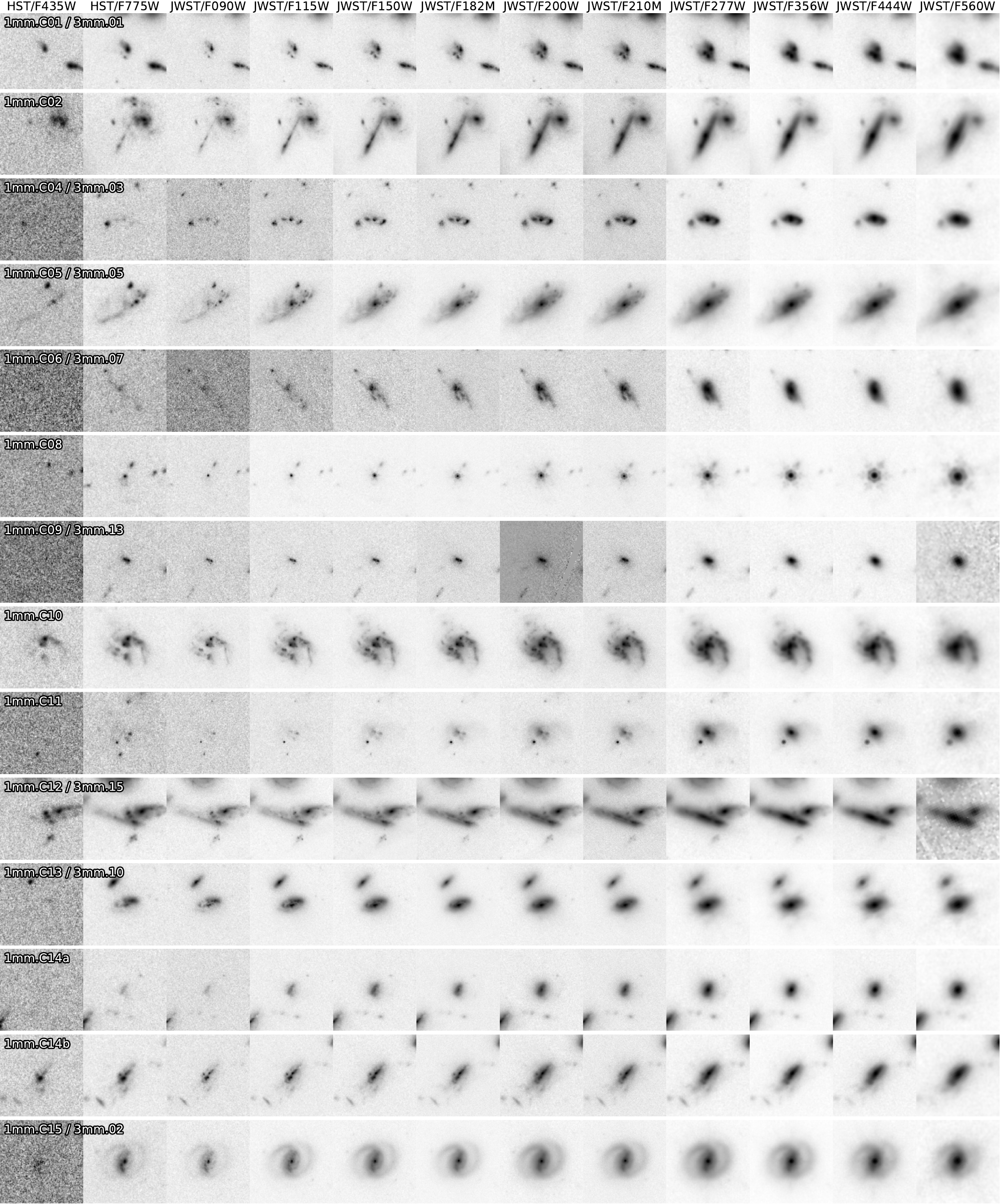}
\caption{Multi-wavelength cutouts for all sources in the sample on the
  same scale as \autoref{fig:cutouts}.
  \label{fig:full-cutouts}}
\end{figure*}

\begin{figure*}[t]
  \figurenum{\ref{fig:full-cutouts}}
  \centering
\includegraphics[width=\textwidth]{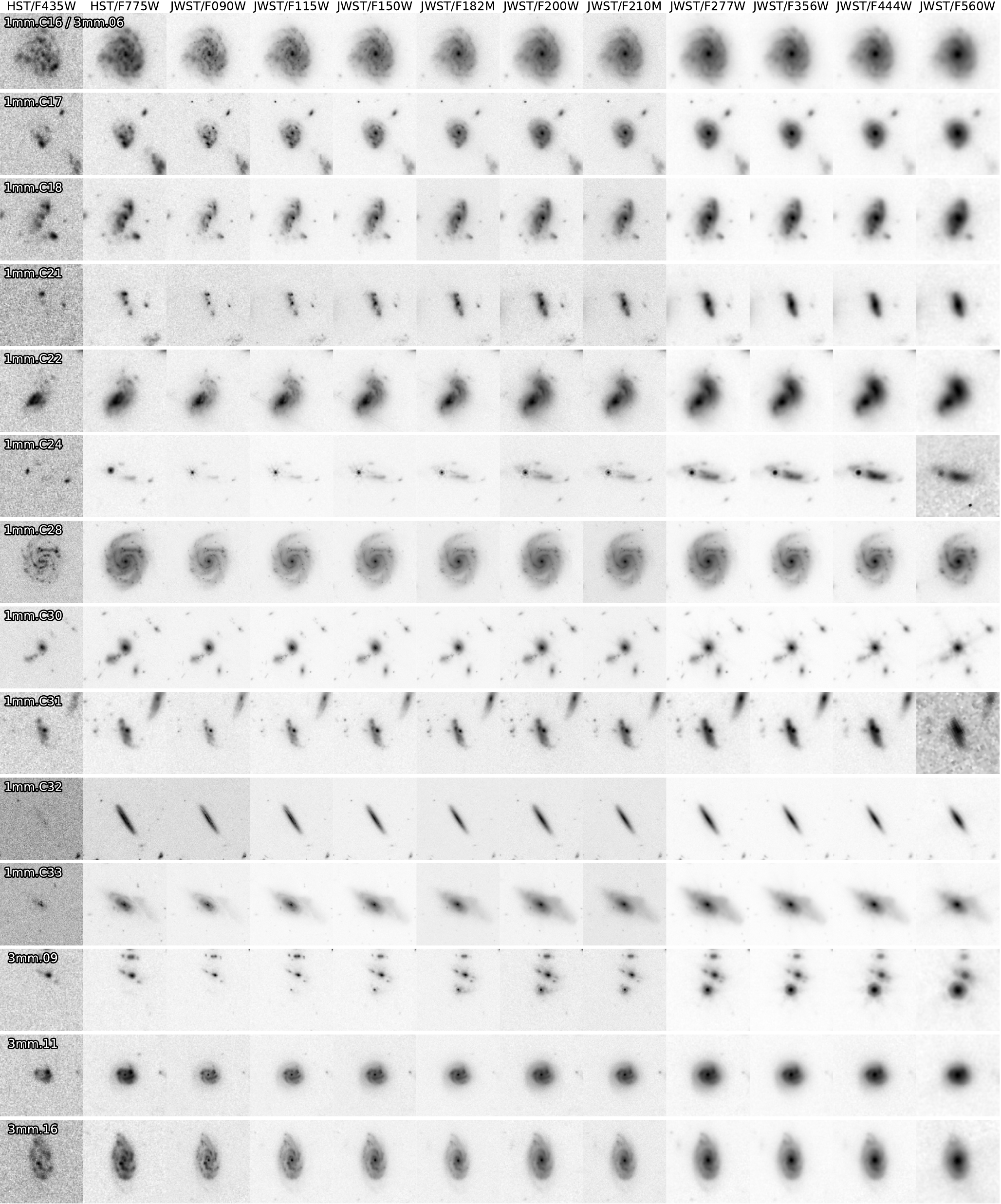}
\caption{\emph{(continued)}}
\end{figure*}

\section{\textsc{galfit} residuals}
\label{sec:galfit-residuals}
We show the \textsc{galfit} residuals for the sources in \autoref{fig:sel-cutouts} in \autoref{fig:galfit-residuals}.

\begin{figure*}[t]
  \centering
\includegraphics[width=\textwidth]{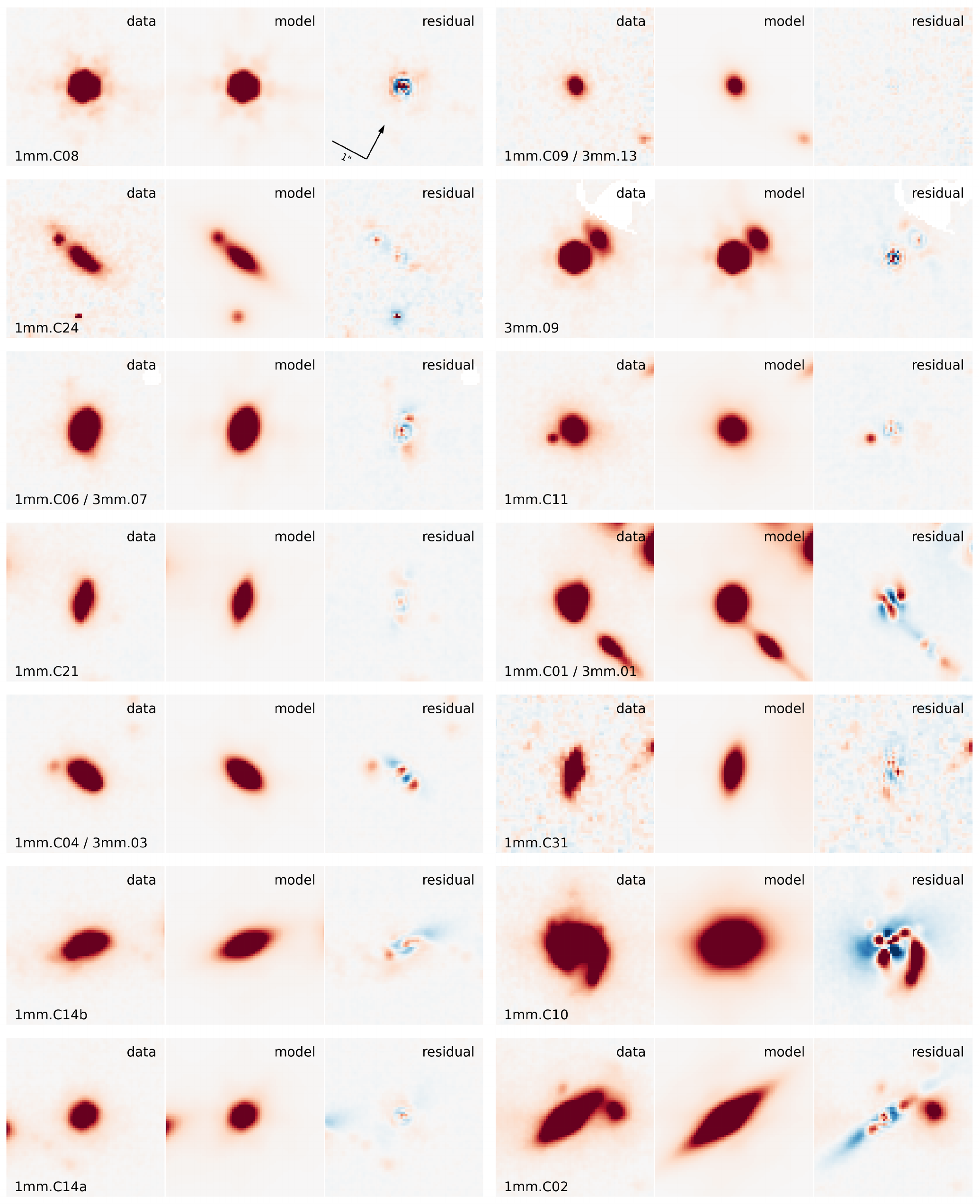}
\caption{\textsc{galfit} modeling results for all sources.  From left
  to right the three panels show the F560W data, the best-fit single
  S\'{e}rsic model, and the residual ($\mathrm{data}-\mathrm{model}$)
  image, all at the same linear scale.  The first ten panels
  correspond to the galaxies from \autoref{fig:sel-cutouts}.  Note
  1mm.C01 has an intrinsic multi-component structure, while for
  1mm.C04 the residuals are driven by the model being affected by the
  second component to the south-west (treating this as one galaxy is
  consistent with earlier work).  These cutouts are the same size as
  those in \autoref{fig:cutouts}, but on the native orientation of the
  MIDIS image at which the fitting is performed, such that north
  points $27.83^{\circ}$ clockwise, as indicated by the 1'' scalebar.
  \label{fig:galfit-residuals}}
\end{figure*}

\begin{figure*}[t]
  \figurenum{\ref{fig:galfit-residuals}}\centering
\includegraphics[width=\textwidth]{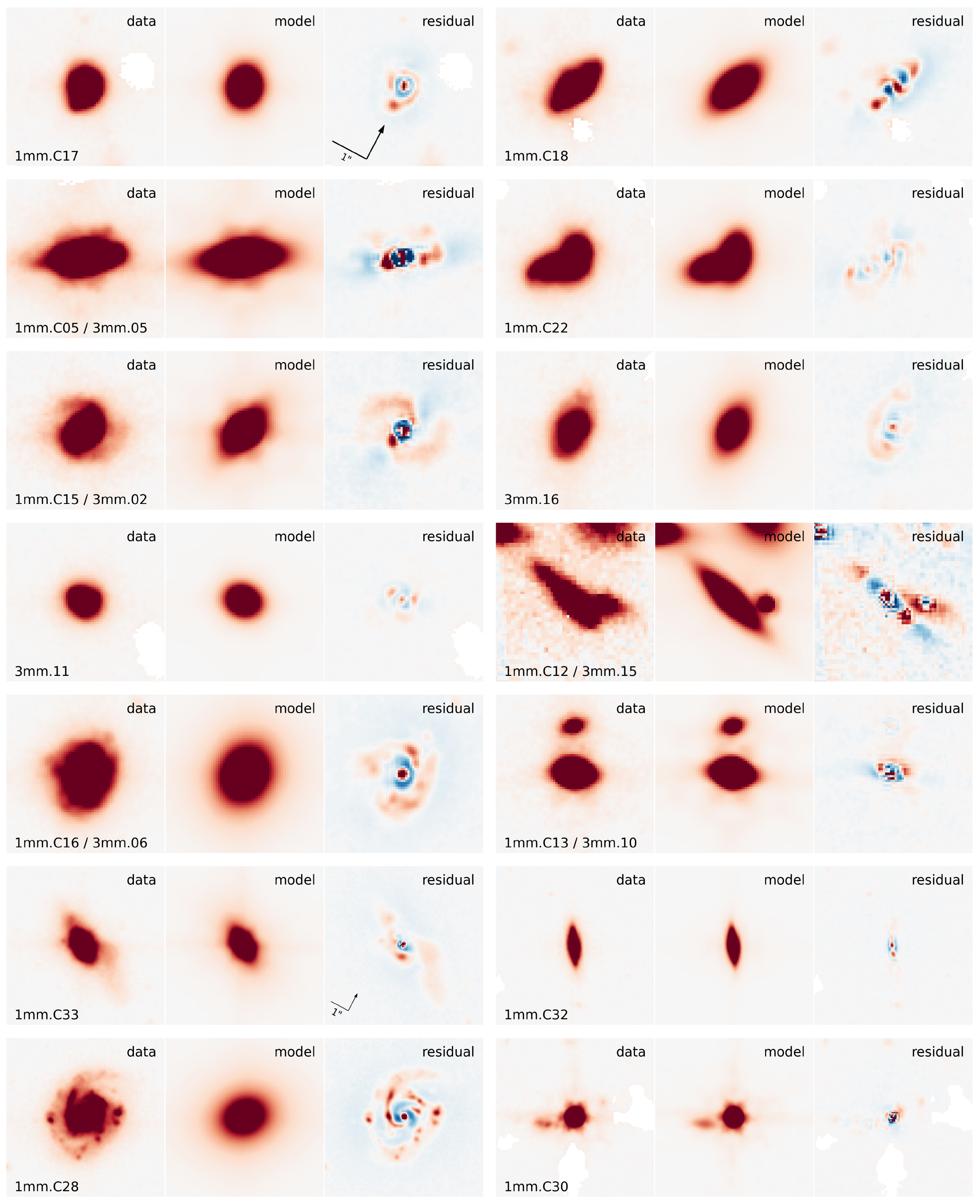}
\caption{\emph{continued}}
\end{figure*}

\end{document}

%% file: sources-final.tex
\begin{deluxetable*}{ccccccccccc}
  \colnumbers \tablecaption{Source properties \label{tab:sources}}
  \tablehead{\colhead{ID 1mm} & \colhead{ID 3mm} & \colhead{R.A.}               & \colhead{Decl.}              & \colhead{$z_{\rm spec}$} & \colhead{Ref} & \colhead{X-ray} & \colhead{$f_{\mathrm{F560W}}$} & \colhead{$M_{*}$}                 & \colhead{$R_{e}$(F560W)} & \colhead{$n$} \\
    \colhead{ } & \colhead{ } & \colhead{$\mathrm{(J2000)}$} &
    \colhead{$\mathrm{(J2000)}$} & \colhead{ } & \colhead{ } &
    \colhead{ } & \colhead{($\mu$Jy)} & \colhead{($10^{10}$
      M$_{\odot}$)} & \colhead{$\mathrm{kpc}$} & \colhead{ } }
  \startdata
1mm.C01       & 3mm.01  & 03:32:38.54  & -27:46:34.6   & 2.543 & CO,M,F & AGN     & 7.790  $\pm$ 0.004 & 1.48  $\pm$ 0.34 & 1.650 $\pm$ 0.083 & 0.643 $\pm$ 0.033 \\
1mm.C02       & \nodata & 03:32:36.96  & -27:47:27.2   & 1.910 & N      & \nodata & 15.112 $\pm$ 0.006 & 8.91  $\pm$ 2.05 & 3.971 $\pm$ 0.199 & 1.589 $\pm$ 0.080 \\
1mm.C03       & 3mm.04  & 03:32:34.45  & -27:46:59.8   & 1.414 & CO,M,F & \nodata & \nodata            & \nodata          & \nodata           & \nodata \\
1mm.C04       & 3mm.03  & 03:32:41.02  & -27:46:31.6   & 2.454 & CO,F   & \nodata & 5.768  $\pm$ 0.004 & 3.98  $\pm$ 0.92 & 2.036 $\pm$ 0.102 & 0.879 $\pm$ 0.045 \\
1mm.C05       & 3mm.05  & 03:32:39.75  & -27:46:11.6   & 1.551 & CO,M   & AGN     & 37.388 $\pm$ 0.008 & 37.15 $\pm$ 8.55 & 2.750 $\pm$ 0.138 & 2.402 $\pm$ 0.120 \\
1mm.C06       & 3mm.07  & 03:32:43.53  & -27:46:39.2   & 2.696 & CO,F   & \nodata & 9.601  $\pm$ 0.008 & 14.13 $\pm$ 3.26 & 1.906 $\pm$ 0.095 & 0.702 $\pm$ 0.035 \\
1mm.C07       & \nodata & 03:32:35.08  & -27:46:47.8   & 2.580 & CO,F   & AGN     & \nodata            & \nodata          & \nodata           & \nodata \\
1mm.C08       & \nodata & 03:32:38.03  & -27:46:26.6   & 3.711 & M,F    & AGN     & 10.875 $\pm$ 0.005 & 30.2  $\pm$ 6.96 & 0.044 $\pm$ 0.006 & 6.077 $\pm$ 0.638 \\
1mm.C09       & 3mm.13  & 03:32:35.56  & -27:47:04.2   & 3.601 & CO,F   & \nodata & 1.562  $\pm$ 0.010 & 0.56  $\pm$ 0.13 & 1.031 $\pm$ 0.054 & 2.610 $\pm$ 0.184 \\
1mm.C10       & \nodata & 03:32:40.07  & -27:47:55.8   & 1.997 & CO,M   & X       & 26.628 $\pm$ 0.010 & 6.31  $\pm$ 1.45 & 2.934 $\pm$ 0.147 & 2.039 $\pm$ 0.103 \\
1mm.C11       & \nodata & 03:32:43.32  & -27:46:47.0   & 2.695 & F      & AGN     & 4.644  $\pm$ 0.005 & 2.63  $\pm$ 0.61 & 2.115 $\pm$ 0.110 & 7.429 $\pm$ 0.391 \\
1mm.C12       & 3mm.15  & 03:32:36.48  & -27:46:31.8   & 1.096 & CO,M,F & AGN     & 12.740 $\pm$ 0.039 & 0.91  $\pm$ 0.21 & 4.855 $\pm$ 0.243 & 1.139 $\pm$ 0.057 \\
1mm.C13       & 3mm.10  & 03:32:42.98  & -27:46:50.2   & 1.037 & CO,M,F & \nodata & 17.113 $\pm$ 0.009 & 16.22 $\pm$ 3.73 & 1.318 $\pm$ 0.066 & 1.393 $\pm$ 0.070 \\
1mm.C14a      & \nodata & 03:32:41.69  & -27:46:55.6   & 1.996 & CO     & \nodata & 4.390  $\pm$ 0.003 & 3.31  $\pm$ 0.76 & 0.997 $\pm$ 0.050 & 5.358 $\pm$ 0.289 \\
1mm.C14b      & \nodata & 03:32:41.846 & -27:46:56.997 & 1.999 & M      & \nodata & 6.290  $\pm$ 0.004 & 1.51  $\pm$ 0.41 & 3.015 $\pm$ 0.151 & 1.374 $\pm$ 0.069 \\
1mm.C15       & 3mm.02  & 03:32:42.37  & -27:47:07.8   & 1.317 & CO,M,F & \nodata & 20.794 $\pm$ 0.014 & 14.45 $\pm$ 3.43 & 1.628 $\pm$ 0.082 & 4.533 $\pm$ 0.229 \\
1mm.C16       & 3mm.06  & 03:32:39.87  & -27:47:15.2   & 1.095 & CO,M,F & X       & 20.214 $\pm$ 0.008 & 5.13  $\pm$ 1.2  & 4.482 $\pm$ 0.224 & 1.590 $\pm$ 0.080 \\
1mm.C17       & \nodata & 03:32:38.80  & -27:47:14.8   & 1.848 & M      & \nodata & 10.631 $\pm$ 0.005 & 4.79  $\pm$ 1.17 & 2.266 $\pm$ 0.113 & 1.268 $\pm$ 0.064 \\
1mm.C18       & \nodata & 03:32:37.36  & -27:46:45.8   & 1.845 & M      & X       & 12.261 $\pm$ 0.006 & 5.01  $\pm$ 1.16 & 3.757 $\pm$ 0.188 & 1.194 $\pm$ 0.060 \\
1mm.C19       & 3mm.12  & 03:32:36.19  & -27:46:28.0   & 2.574 & CO,M,F & AGN     & \nodata            & \nodata          & \nodata           & \nodata \\
1mm.C20       & \nodata & 03:32:35.77  & -27:46:27.6   & 1.093 & M,F    & \nodata & \nodata            & \nodata          & \nodata           & \nodata \\
1mm.C21       & \nodata & 03:32:35.98  & -27:47:25.8   & 2.643 & F      & \nodata & 3.830  $\pm$ 0.004 & 2.29  $\pm$ 0.53 & 2.317 $\pm$ 0.116 & 0.848 $\pm$ 0.043 \\
1mm.C22       & \nodata & 03:32:37.60  & -27:47:44.2   & 1.542 & M,F    & \nodata & 8.786  $\pm$ 0.006 & 4.79  $\pm$ 1.1  & 2.987 $\pm$ 0.149 & 0.798 $\pm$ 0.040 \\
1mm.C23       & 3mm.08  & 03:32:35.55  & -27:46:26.2   & 1.382 & CO,M,F & \nodata & \nodata            & \nodata          & \nodata           & \nodata \\
1mm.C24       & \nodata & 03:32:38.76  & -27:48:10.4   & 2.823 & F      & \nodata & 3.136  $\pm$ 0.018 & 1.58  $\pm$ 0.36 & 3.255 $\pm$ 0.167 & 1.756 $\pm$ 0.094 \\
1mm.C25       & 3mm.14  & 03:32:34.85  & -27:46:40.6   & 1.098 & CO,M,F & \nodata & \nodata            & \nodata          & \nodata           & \nodata \\
1mm.C26       & \nodata & 03:32:34.70  & -27:46:45.0   & 1.552 & M,F    & \nodata & \nodata            & \nodata          & \nodata           & \nodata \\
1mm.C28       & \nodata & 03:32:40.85  & -27:46:16.4   & 0.622 & M      & \nodata & \nodata            & \nodata          & 8.955 $\pm$ 0.453 & 1.686 $\pm$ 0.085 \\
1mm.C30       & \nodata & 03:32:38.79  & -27:47:32.4   & 0.458 & M,F    & X       & 28.474 $\pm$ 0.007 & 0.19  $\pm$ 0.04 & 0.559 $\pm$ 0.028 & 2.287 $\pm$ 0.115 \\
1mm.C31       & \nodata & 03:32:37.08  & -27:46:17.4   & 2.227 & M,F    & \nodata & 4.154  $\pm$ 0.027 & 0.74  $\pm$ 0.2  & 3.066 $\pm$ 0.155 & 0.759 $\pm$ 0.044 \\
1mm.C32       & \nodata & 03:32:37.75  & -27:47:06.8   & 0.667 & M      & \nodata & 11.310 $\pm$ 0.005 & 1.74  $\pm$ 0.4  & 3.471 $\pm$ 0.174 & 0.814 $\pm$ 0.041 \\
1mm.C33       & \nodata & 03:32:38.50  & -27:47:02.8   & 0.948 & M      & \nodata & 32.469 $\pm$ 0.008 & 13.8  $\pm$ 3.21 & 3.974 $\pm$ 0.199 & 5.240 $\pm$ 0.263 \\
UDF-1         & 3mm.09  & 03:32:44.03  & -27:46:36.0   & 2.698 & CO,F   & AGN     & 11.329 $\pm$ 0.006 & 23.44 $\pm$ 5.4  & 0.224 $\pm$ 0.012 & 3.279 $\pm$ 0.179 \\
\nodata       & 3mm.11  & 03:32:39.81  & -27:46:53.5   & 1.096 & CO,M,F & \nodata & 6.210  $\pm$ 0.005 & 1.55  $\pm$ 0.36 & 2.430 $\pm$ 0.122 & 0.853 $\pm$ 0.043 \\
Faint.1mm.C20 & 3mm.16  & 03:32:39.92  & -27:46:07.4   & 1.294 & CO,M,F & \nodata & 9.020  $\pm$ 0.008 & 3.09  $\pm$ 0.73 & 3.915 $\pm$ 0.196 & 2.081 $\pm$ 0.104 
  \enddata \tablecomments{(1) ALMA 1mm Continuum ID (1mm.C\#\#:
    ASPECS-LP \citealt{Gonzalez-Lopez2020, Aravena2020}; UDF\#:
    \citealt{Dunlop2017}) (2) ALMA 3mm CO ID
    \citep{Gonzalez-Lopez2019, Boogaard2019} (3) Right Ascension (4)
    Declination (5) Redshift (6) Redshift reference: CO: CO and/or
    \CI\ line(s) \citep{Boogaard2019, Boogaard2020}, M: MUSE
    \citep[][]{Inami2017, Boogaard2019, Bacon2023}, F: FRESCO
    \citep{Oesch2023}, N: NGDEEP \citep{Bagley2023b,Pirzkal2023}.  (7)
    AGN or other X-ray Source \citep{Luo2017}. (8) MIRI/F560W flux
    density from MIDIS.  (9) Stellar mass.  (10) Effective radius.
    (11) S\'{e}rsic index.}
\end{deluxetable*}